%% file: bare_conf_compsoc.tex
\documentclass[conference,compsoc]{IEEEtran}

\usepackage{enumitem}
\usepackage{subcaption}
\usepackage{multirow}
\usepackage{booktabs}
\usepackage{tabularx}
\usepackage{xspace} %
\usepackage{ragged2e}
\usepackage{makecell}
\usepackage{xcolor}
\usepackage{xurl}
\usepackage{framed}

\ifCLASSOPTIONcompsoc
  \usepackage[nocompress]{cite}
\else
  \usepackage{cite}
\fi

\ifCLASSINFOpdf
  \usepackage[pdftex]{graphicx}
\else
\fi

\begin{document}

\newcommand{\sssec}[1]{\vspace*{0.05in}\noindent\textbf{#1}}
\newcommand{\name}{Privacy Detective\xspace}
\newcommand{\todo}[1]{\textcolor{blue}{#1}}
\newcommand{\revision}[1]{\textcolor{black}{#1}}
\newcommand{\haojian}[1]{\textcolor{red}{#1}}
\title{\textit{\name}: A Narrative Game that Cultivates Student Developers' Privacy Awareness by Harnessing Legal Documents}

\author{
    \IEEEauthorblockN{
        Shao-Yu Chu\IEEEauthorrefmark{2}, 
        Jennifer Forsyth\IEEEauthorrefmark{2}, 
        Xu Wang\IEEEauthorrefmark{4}, 
        Haojian Jin\IEEEauthorrefmark{2},
    \IEEEauthorblockA{
        \IEEEauthorrefmark{2}University of California, San Diego
    \\\{shaoyuchu, jcforsyth, haojian\}@ucsd.edu}
    \IEEEauthorblockA{\IEEEauthorrefmark{4}University of Michigan
    \\\{xwanghci\}@umich.edu}
    }
}

\maketitle

\input{parts/p0-abstract}

\IEEEpeerreviewmaketitle

\input{parts/p1-intro}

\input{parts/p2-related}

\input{parts/p3-design}

\input{parts/p4-harness-legal-documents}

\input{parts/p5-design-iteration}

\input{parts/p6-evaluation}

\input{parts/p7-discussion}

\input{parts/p10-ethical-considerations}

\input{parts/p11-conclusion}

\newpage

\bibliographystyle{IEEEtran}
\bibliography{sample-base}

\input{parts/p99-appendix}

\end{document}

%% file: parts/p0-abstract.tex
\begin{abstract}

Developers’ choices about what data a system collects, how it is used and shared, and what defaults govern user choices directly shape users’ privacy experiences. Yet, developers often make problematic privacy-related design decisions without realizing the potential consequences. 
We introduce \textit{\name}, a narrative investigation game that leverages real-world legal documents to train developers' privacy awareness. 
In the game, players search for privacy violation evidence derived from legal documents and organize this evidences into privacy violation reports using curated templates. 
We evaluated \name in a between-subjects study with student developers, comparing it against a baseline in which participants read raw FTC legal documents. Participants in the game condition identified more true violations than the baseline group, flagged fewer non-issues, and provided more complete justifications for the violations they reported.

\end{abstract}

%% file: parts/p1-intro.tex
\section{Introduction}

Software developers play a central role in shaping how systems handle users' personal data~\cite{jain2014should}. They decide which libraries, tools, and platforms to use, what data to collect, and how to present information to users. These choices directly impact user privacy~\cite{jain2014should, tahaei2022embedding, tahaei2021privacy}: how personal data flows, and whether users retain agency over their own information and choices.
However, developers frequently make problematic privacy-related design decisions without realizing the potential consequences~\cite{rubinstein2013privacy}. Even when privacy experts are available on a team, low privacy awareness means developers may not recognize privacy-relevant issues in the first place, preventing them from flagging or escalating concerns to experts~\cite{horstmann2024those, horstmann2025sorry}.

In response, many universities have introduced privacy-focused courses into computer science curricula~\cite{securecomputersystemcourse,shilton2020role,tahaei2019don,moore2018introducing}. Companies likewise emphasize privacy in onboarding and refresher trainings for engineers~\cite{prybylo2024evaluating}. These trainings typically rely on two instructional formats: (1) abstract, concept-based quizzes (e.g., ``Which of the following best describes personally identifiable information?'') and (2) simple scenario-based multiple-choice questions (e.g., ``What action would you take in situation X?'')~\cite{android_privacy_security_quiz, quizlet_hipaa_privacy_flashcards}. However, a major concern is that these materials often remain disconnected from real-world development practices~\cite{tahaei2021privacy}. 
Recent studies found that developers more commonly cultivate sensitivity to real-world privacy problems informally--by reading online resources such as technical articles, online forums, news coverage, general media, and blogs about real incidents~\cite{tahaei2021privacy, prybylo2024evaluating}. Since privacy habits and awareness can begin to form early in education, student developers represent one critical target population for more effective privacy training.

In this paper, we present \name, a narrative investigation game that leverages real-world legal documents to actively train \revision{student} developers' privacy awareness. 
\name aims to sharpen their sensitivity to potential privacy risks when they examine real-world data practices. If developers can recognize when ``something feels off,'' they may escalate the issue to privacy experts, who can help navigate the nuances of implementation~\cite{prybylo2024evaluating}.

Recognizing when ``something feels off'' is not easy. First, modern software involves countless interconnected design decisions, so developers must filter out privacy-irrelevant details to focus their attention on the ones that matter. This contrasts with conventional training, where developers are given carefully selected multiple-choice scenarios in which the relevant decision points and options are already highlighted. Second, many developers possess limited knowledge of established privacy principles and lack the reasoning skills necessary to apply them in real-world contexts. Third, privacy sensitivity is inherently subjective and exists on a spectrum, making it difficult to determine whether developers are being overly cautious or insufficiently attentive to privacy risks.

\begin{figure*}[h]
    \centering
    \includegraphics[width=0.9\linewidth]{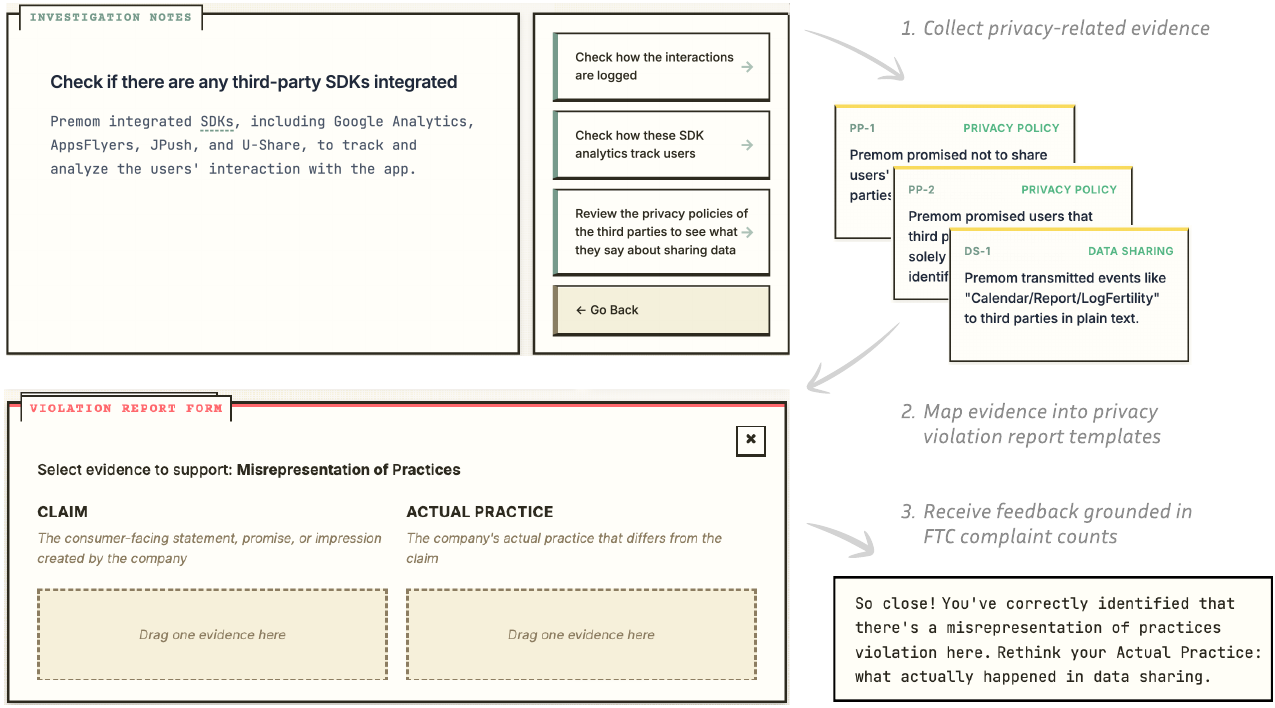}
    \caption{In \name, players investigate a case by searching for evidence grounded in legal documents, then assemble that evidence into privacy violation reports using curated templates.}
    \label{fig:screenshot}
\end{figure*}

\name helps developers overcome these barriers by allowing them to practice their privacy examination skills in a narrative game.
Players in \name investigate detailed descriptions of a data practice, collect privacy-related evidence, and organize the collected evidence into privacy violation reports. 
The key difference between \name and a conventional scenario-based question is that it requires developers to actively search for evidence in textual descriptions and reason about the privacy violations based on collected evidence.

\name does not require developers to master privacy principles in advance. Instead, it encodes those principles in a structured data schema and provides a privacy violation report template that guides players to slot relevant evidence into predefined fields. For example, a Misrepresentation of Practices violation~\cite{unfairftc} can be represented with two fields: Claim (the consumer-facing statement, promise, or impression created by the company) and Actual Practice (what the company actually does that conflicts with that claim). By repeatedly mapping concrete evidence to these structured fields, developers gradually internalize the underlying privacy principles and learn how to apply them in real-world contexts.

\name derives its narrative game scenarios from enforcement-related documents published by the U.S. Federal Trade Commission (FTC), which cover a broad range of practices that compromise users' privacy and autonomy. These documents are informative: they detail the data practices at issue, enumerate the specific legal violations alleged (counts), and explain the rationale behind each count. \name uses the data-practice descriptions to construct the searchable environment and in-game scenes, treats the counts as the ground-truth set of applicable violations, and uses the accompanying rationale to provide targeted feedback when players submit reports.

We conducted a between-subjects study (N=36) comparing \name to a reading-based approach in terms of cultivating privacy awareness. In the reading condition, participants learned from FTC cases by skimming the official press release and complaint under the same time limits as the game, mirroring how developers typically use news and online articles to learn about privacy issues~\cite{tahaei2021privacy, prybylo2024evaluating}. Our results showed that \name significantly improved participants' ability to flag true violations for escalation (+0.24 recall) without increasing over-flagging (+0.13 precision) and to justify violations more completely (+0.22 completeness; i.e., the ratio of true positives that included the correct violation type and all key evidence). 
The reading condition showed smaller gains across all three metrics, with a recall gain less than half the size of the game's ($\beta_{Reading}=0.099$ vs. $\beta_{Game}=0.242$), and no significant improvements in precision or reasoning completeness.
We release \name as a publicly accessible web artifact at \textcolor{blue}{\url{https://privacy-detective.vercel.app/}}.

\textbf{Contributions}. 
This paper makes three contributions. First, we introduce \name, a novel class of training questions for developers' privacy education.
Second, through a controlled empirical study, we provide evidence that this detective game approach sharpens student developers' sensitivity to potential privacy risks in real-world data practices, improving their ability to detect, justify, and reason about privacy violations.
Third,  \name demonstrates a novel way to leverage the growing corpus of legal enforcement documents, using these documents to help developers calibrate their privacy judgments against concrete regulatory outcomes rather than abstract principles.

%% file: parts/p2-related.tex
\section{Related Work}

This project builds on ideas from three key areas: (1) developer privacy awareness, (2) narrative games as learning tools, and (3) learning through legal documents.

\subsection{Developer Privacy Awareness}

Awareness is a prerequisite for behavior change~\cite{ludwig2020self,das2016social} --- for both developers and end users. Limited privacy awareness can prevent end users from recognizing when a product collects, uses, or shares personal data, and from taking protective actions such as adjusting settings, withholding information, or opting out~\cite{bergmann2008testing, malandrino2013increased, potzsch2008privacy, yao2019designing}. Researchers have developed various interventions to cultivate users' privacy awareness, including privacy nudges (e.g., just-in-time notifications~\cite{almuhimedi2015your, balebako2013little} and presentation framing~\cite{adjerid2013sleights, acquisti2017nudges, feng2021design}), visual aids (e.g., privacy dashboards~\cite{farke2021privacy, thakkar2022would, herder2020privacy} and privacy nutrition labels~\cite{kelley2009nutrition, li2022understanding, kelley2010standardizing}), and educational programs~\cite{desimpelaere2020knowledge, gerber2018foxit}. 

Our work focuses on strengthening developers’ privacy awareness, because developers’ choices about what data a system collects, how it is used and shared, and what defaults govern user choices directly shape users’ privacy experiences~\cite{tahaei2022embedding, rubinstein2013privacy}. 
Developers lacking privacy awareness may unintentionally expand collection, sharing, and retention, increasing risks to users~\cite{Wired2021PowerApps, Riley2020Premom, bbcUberGodView2016} and creating legal and reputational liabilities for organizations~\cite{FTC2019Facebook, FTC2023AmazonAlexa, FTC2021FloPrivacy}. For example, in the Facebook--Cambridge Analytica incident~\cite{isaak2018user, FTC2019Facebook}, millions of users had their personal data harvested for political profiling without consent~\cite{Cadwalladr2018CambridgeAnalytica}. Facebook consequently faced multi-billion-dollar fines, lawsuits, and lasting reputational damage~\cite{FTC2019Facebook}.

Researchers have explored several educational methods to cultivate developers’ privacy awareness. These include university curricula~\cite{boteju2025know, moore2018introducing}, certificate programs for professionals~\cite{iapp_cipt_training}, and game-based interventions that teach privacy awareness through engaging activities~\cite{tian2025panopticon, arachchilage2020designing, alhazmi2022developers, shilton2020role}. For example, Tian~\cite{tian2025panopticon} designed an educational board game that helps students learn privacy-sensitive data practices through hands-on design activities and structured peer critique.
Unlike prior educational efforts, our approach grounds privacy awareness training in real-world enforcement cases.

\subsection{Narrative Games as Learning Tools}

Games can be effective learning tools, as they engage learners through intrinsic motivation while enabling iterative, low-stakes practice with immediate feedback~\cite{plass2015foundations, tobias2013game, krath2021revealing}. Narrative games extend these affordances by embedding gameplay within a story structure, where players assume a role in an unfolding plot and take actions such as investigation, dialogue, and consequential choices~\cite{pilegard2016improving, dubbelman2016narrative}. They scaffold learning by organizing complex scenarios into coherent sequences that reduce cognitive load~\cite{naul2020story, dickey2006game}. By placing learners in a role, narrative games also support situated learning~\cite{anderson1996situated, young1993instructional}, where learners develop knowledge through participating in authentic activities within a realistic context. This helps learners reason from contextual cues and constraints and practice judgment rather than applying abstract rules in isolation~\cite{dickey202011, ammanabrolu2021situated, gee2003video, lin2022learning}. Researchers have applied narrative games to learning across domains, including STEM~\cite{rowe2013embedded, pilegard2016improving}, reasoning skills (e.g., critical thinking)~\cite{fallon2013making}, language learning~\cite{chen2018using}, and socio-technical topics such as law~\cite{moshirnia2020ludic}, ethics~\cite{maheshwary2025case, grasse2022using}, and cybersecurity~\cite{kc7cyber}. Building on prior narrative-based educational games, we apply narrative investigation mechanics to privacy education.

\subsection{Learning through Legal Documents}

Legal education has long relied on case-based learning~\cite{thistlethwaite2012effectiveness, kantar2015case} to teach judgment from precedent~\cite{williams1992putting, weaver1991langdell, patterson1951case}. Case-based learning centers instruction on contextualized cases from the real-world~\cite{williams2005case}, helping learners connect abstract principles to concrete situations they might encounter in the workspace~\cite{bransford2000people}. It is especially effective in ill-structured domains where problem boundaries are ambiguous and sound decisions depend on interpretation and justification rather than fixed procedures~\cite{spiro2019cognitive, jonassen1997instructional}.

Beyond training lawyers, case-based learning can similarly help learners in socio-technical domains~\cite{martin2021using, ahmad2021case}. For example, Tammeleht et al. trained both undergraduate and graduate students in research ethics and integrity through collaborative workshops where groups analyzed ethical dilemma cases~\cite{tammeleht2019collaborative}. Our work extends case-based learning to privacy awareness by transforming real enforcement documents into an interactive narrative game.

%% file: parts/p3-design.tex
\section{An Overview of \name}

This section describes the key design decisions behind \name and explains how each decision supports our goal of training developers’ privacy awareness. 

\subsection{Motivation and Design Goal}

This project is motivated by findings reported in prior research studies~\cite{tahaei2021privacy,senarath2018developers,shilton2019linking} and by our conversations with multiple privacy practitioners in industry, which reveal a critical gap in developer privacy training. Many companies have adopted a three-layer organizational structure to manage privacy risks, comprising privacy experts, privacy champions, and regular developers~\cite{tahaei2021privacy}. However, a key challenge for them is the scarcity of effective resources to train regular developers to become privacy champions—or to equip them with basic privacy awareness.

Conventional training materials, such as multiple-choice questions about the definition of data minimization principles, demonstrate limited effectiveness because they lack grounding in real-world development scenarios. As noted by privacy practitioners in prior research~\cite{tahaei2021privacy}, developers struggle to translate abstract privacy concepts into practical knowledge when training is disconnected from their actual technical work.
Similarly, Shilton et al. found that developers gain ethical and privacy reasoning skills through exposure to concrete cases, reflection, and comparison—rather than through checklists alone~\cite{shilton2019linking}.

Therefore, we are motivated to develop contextually relevant, scenario-based training that bridges this gap.
Our specific focus is to actively cultivate developers’ sensitivity to potential privacy risks as they examine real-world data practices—helping them recognize when “something feels off”. By strengthening this early detection skill, we hope that developers can flag questionable practices for further review so that Privacy Champions and other privacy specialists can step in, provide guidance, and help navigate the nuanced implementation decisions that often determine whether a system’s data handling is appropriately privacy-protective.

\begin{figure*}[h!]
	\centering
	\begin{subfigure}{0.46\textwidth}
	\centering
	\includegraphics[width = \textwidth]{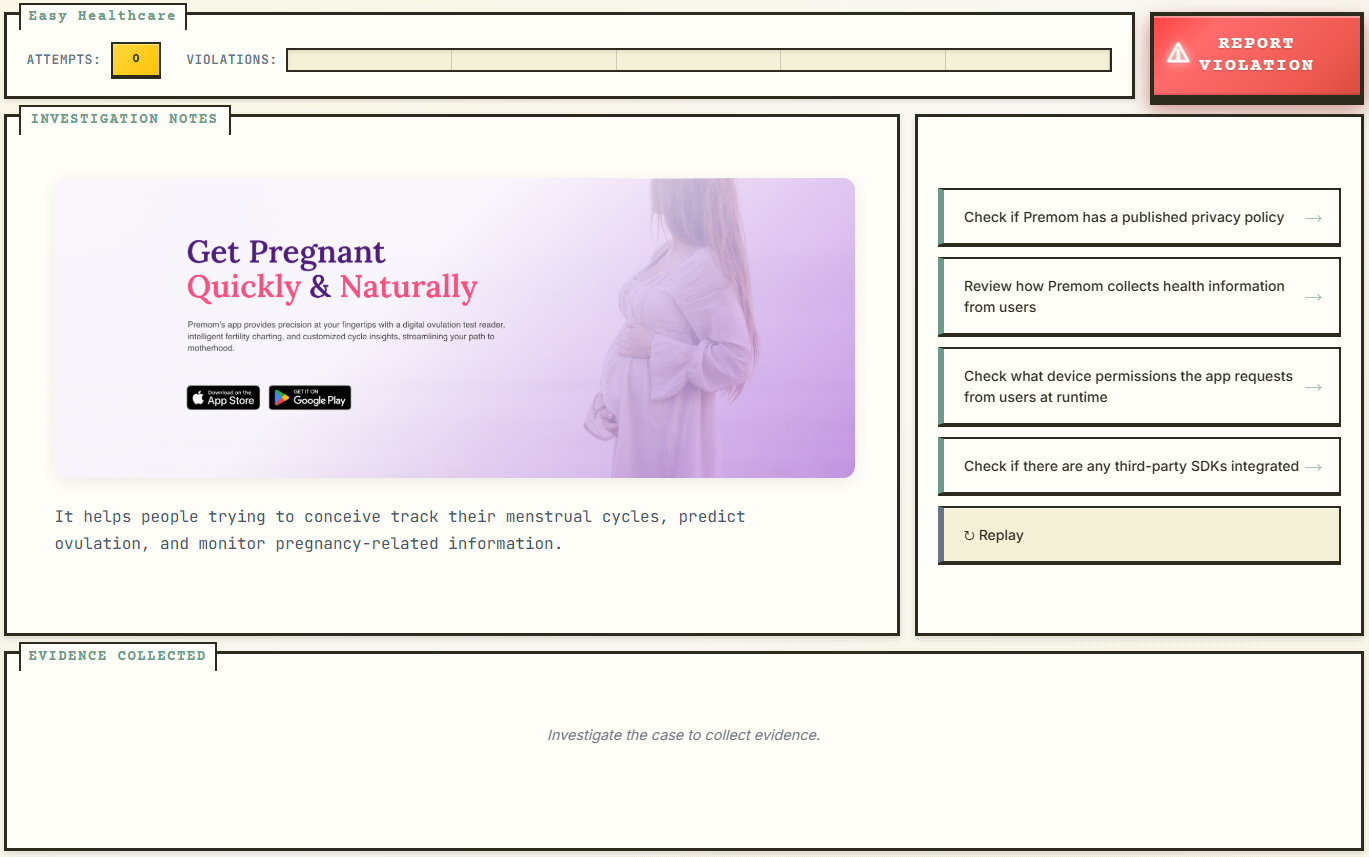}
	\caption{Explore facts in real practices}
	\label{fig:left}
	\end{subfigure}
         \hfill
	\begin{subfigure}{0.49\textwidth}
	\centering
	\includegraphics[width = \textwidth]{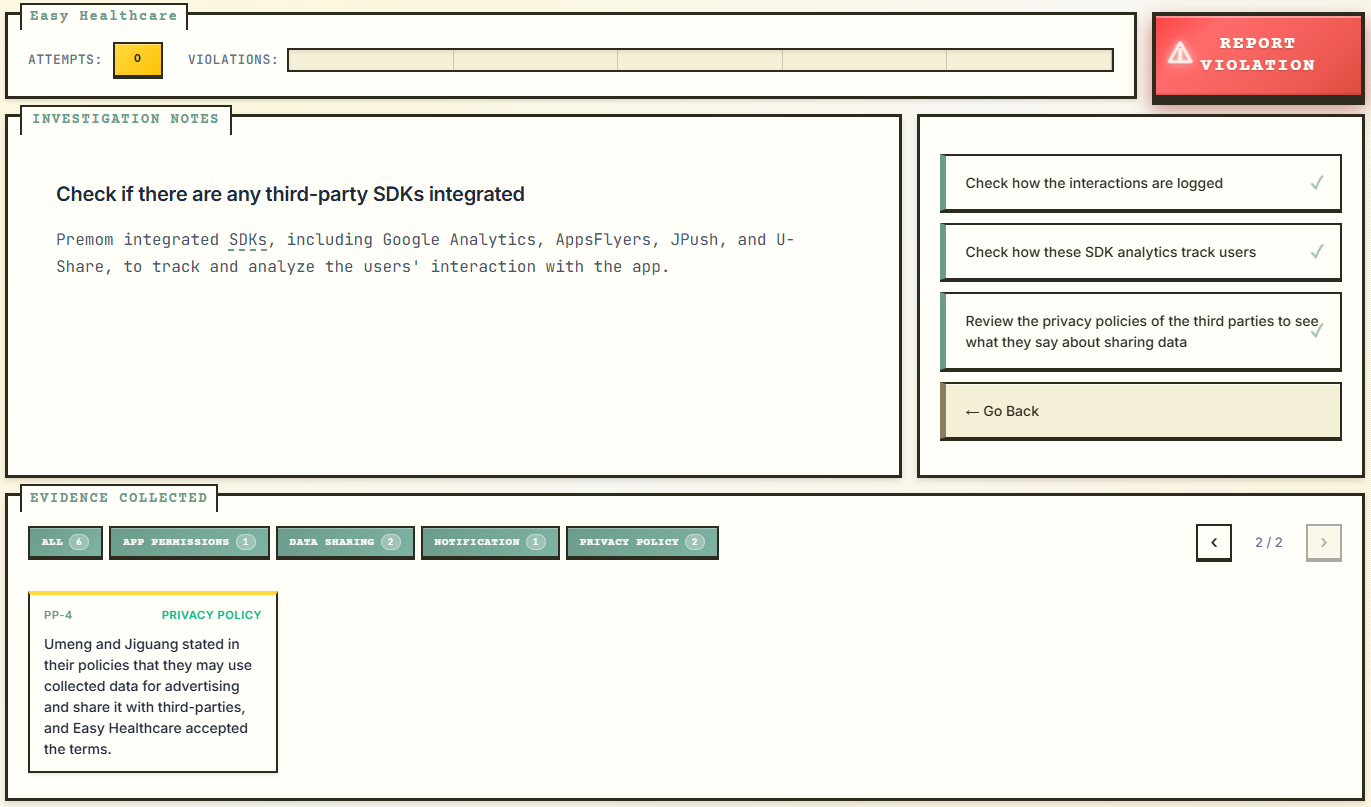}
	\caption{Bookkeep evidence automatically}
	\label{fig:right}
	\end{subfigure}
    \hfill
	\begin{subfigure}{0.49\textwidth}
	\centering
	\includegraphics[width = \textwidth]{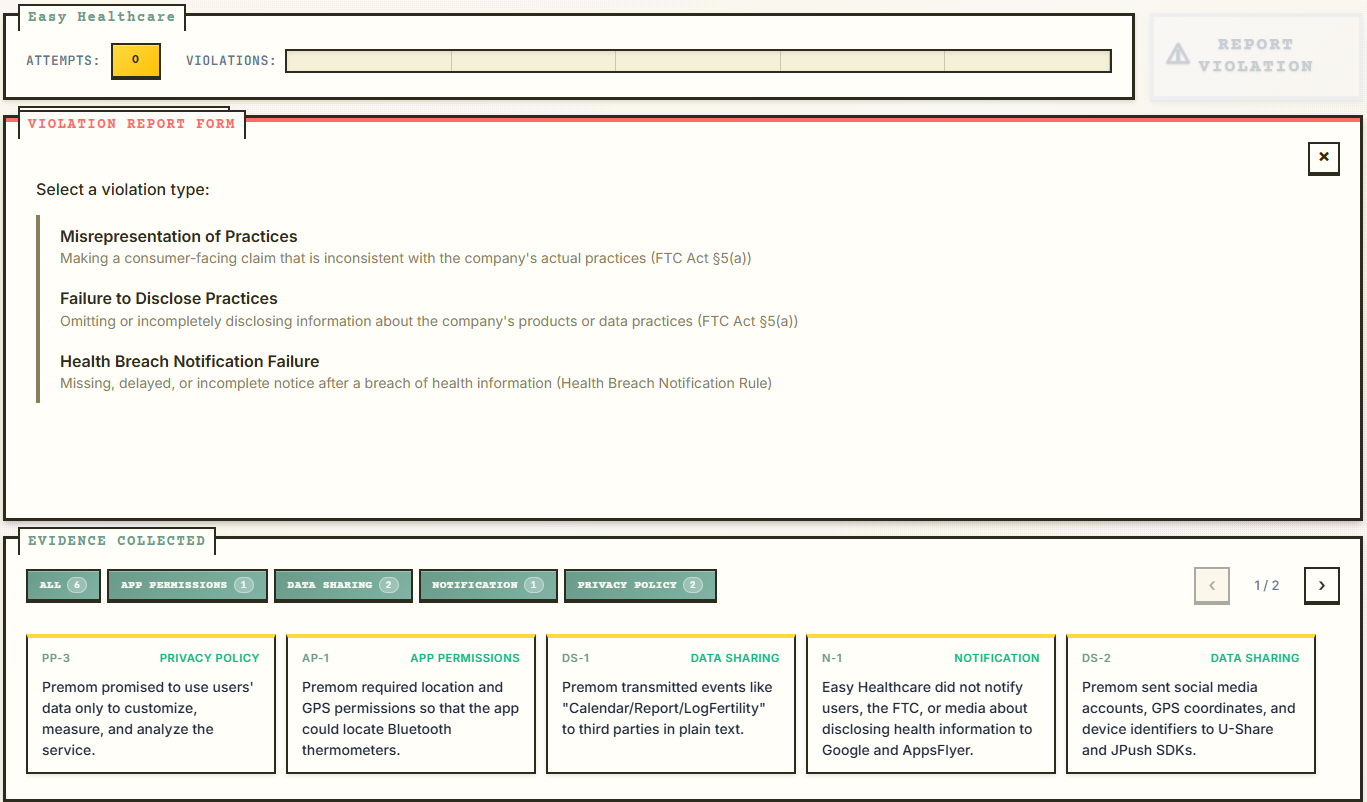}
	\caption{Select the violation type}
	\label{fig:center}
	\end{subfigure}
\hfill
	\begin{subfigure}{0.49\textwidth}
	\centering
	\includegraphics[width = \textwidth]{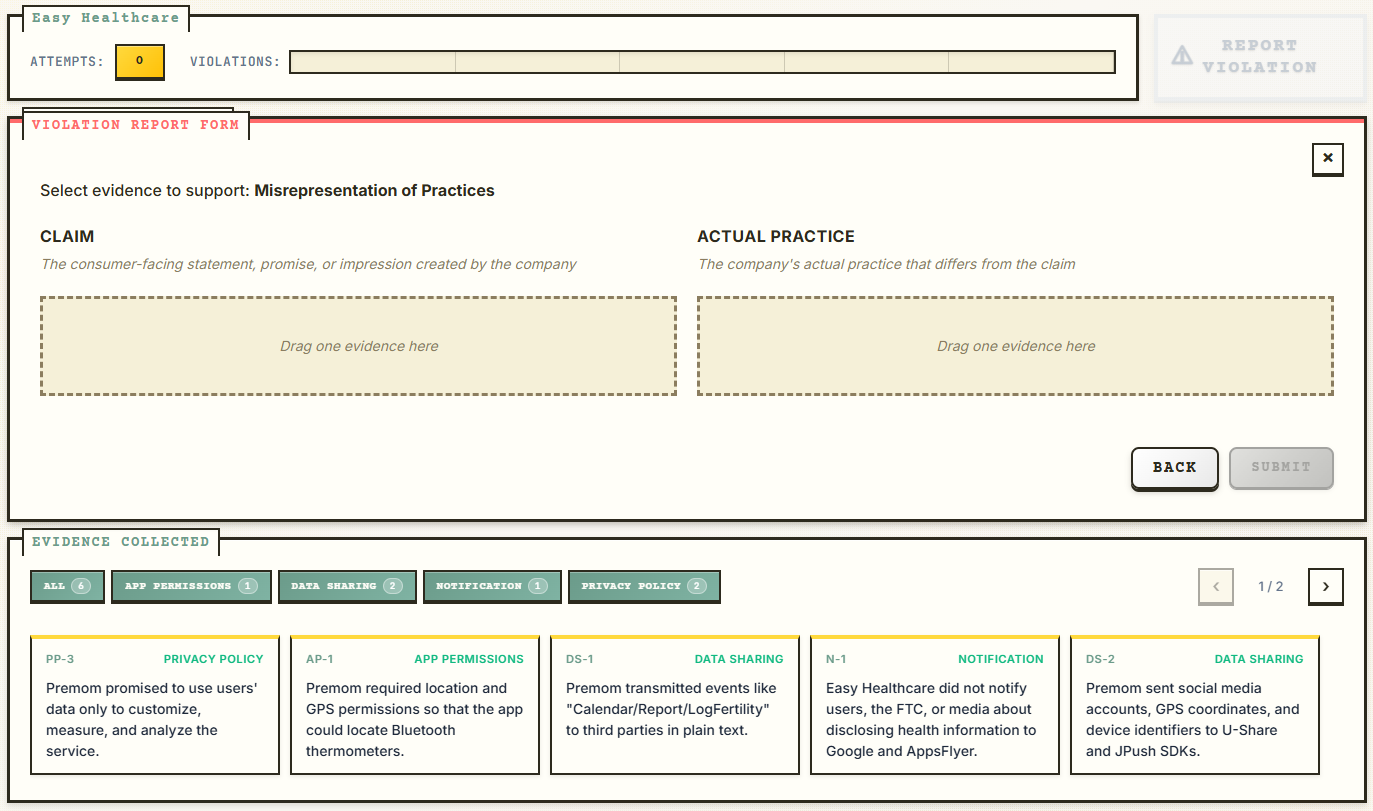}
	\caption{Map evidence to violation templates}
	\label{fig:right}
	\end{subfigure}
	\caption{In-game screenshots. The scenario is adapted from an FTC settlement with Easy Healthcare, which alleged that the company misled users by disclosing sensitive personal information to third parties~\cite{EasyHeal44:online}.}
	\label{fig:screenshots}
\end{figure*}

\subsection{Professional Vision for Privacy}
\label{professional-vision-for-privacy}

We ground our approach in the concept of professional vision, which describes how experts learn to “see” the world in ways that are shaped by their training and practice~\cite{goodwin2015professional,marmor2016vision}. Prior work shows that experts do not simply possess more knowledge than novices; rather, they attend to selective aspects of a situation and interpret them through domain-specific frameworks. For example, a trained painter viewing a scene immediately identifies which background elements can be simplified to foreground the subject, whereas a novice tends to treat all visual elements as equally salient and renders them uniformly~\cite{marmor2016vision,chi1981categorization}. This expert way of seeing reflects learned patterns of attention and interpretation, not merely technical skill. Similarly, when examining a digital product or service, privacy experts look beyond surface features or policy text to perceive underlying information flows, implicit purposes of data collection, and potential points of misuse or harm. They allocate their limited attention to subtle cues of opacity, over-collection, or ambiguous consent and connect these concrete design choices to abstract privacy principles such as transparency, purpose limitation, and user agency—connections that non-experts often fail to make.

Viewing privacy sensitivity through the lens of professional vision also clarifies what, concretely, can be trained. Rather than treating sensitivity as a single construct, we decompose it into three interlocking sub-skills: (1) selective attention, (2) privacy knowledge and reasoning, and (3) privacy sensitivity calibration. Selective attention is the ability to prioritize privacy-relevant aspects of a system despite many competing concerns. Privacy knowledge and reasoning enable developers to explain why a noticed practice is problematic by connecting concrete design decisions to privacy concepts and principles. Privacy sensitivity calibration tunes one’s threshold for concern, preventing both missing meaningful issues and over-flagging benign practices.

\subsection{Game Mechanics}

Designing an engaging and pedagogically effective game from the ground up involves numerous interdependent design decisions.
Rather than inventing game mechanics from scratch, we grounded our design in narrative investigation games~\cite{dubbelman2016narrative} -- a genre that embeds gameplay within a story structure in which players assume a role and take actions such as investigation. This genre maps naturally onto the three sub-skills we aim to cultivate: the investigative structure requires players to exercise selective attention to identify what matters, apply privacy knowledge to reason about what they find, and calibrate judgment against concrete outcomes. Prior work has applied narrative investigation mechanics across educational domains including legal education~\cite{moshirnia2020ludic}, critical thinking~\cite{fallon2013making}, and policy reasoning~\cite{easterday2012policy}, supporting this genre as a learning tool.

We incorporated four design features that operationalize the training of our three target sub-skills (Table~\ref{table:gamefeatures}):

\begin{table}[h]
\centering
\small
\caption{Mapping of game features to learning objectives.}
\begin{tabularx}{\linewidth}{p{0.24\linewidth} X p{0.28\linewidth}}
\toprule
\textbf{Learning Objective} & \textbf{Description} & \textbf{Game Features} \\
\midrule
\textbf{Selective Attention} & 
Ability to focus on the most relevant elements in a complex scenario. &
Situated progressive search \\

\addlinespace
\textbf{Privacy Knowledge \& Reasoning} &
Understanding of core privacy principles and applying them on real-world contexts. &
Templated reasoning, Feedback on report submissions \\

\addlinespace
\textbf{Privacy Sensitivity Calibration} &
Setting an appropriate threshold for concern, avoiding over-caution and inattention. &
FTC counts as ground truth \\
\bottomrule
\end{tabularx}
\label{table:gamefeatures}
\end{table}

\sssec{Situated Progressive Search.} Developing selective attention requires that learners practice identifying privacy-relevant signals while filtering out irrelevant information. Conventional multiple-choice questions and scenario quizzes bypass the challenge by presenting only privacy-relevant features to learners. In practice, developers reviewing a real system encounter privacy-relevant decisions embedded within a much larger set of privacy-neutral ones. At the same time, presenting learners with the full complexity of a commercial software system, including its codebase and user-facing policies, creates a competing problem. Without structure, the information load overwhelms the learners before they can reason about what they have found. Effective training should preserve the discrimination task while keeping cognitive load tractable.

We address this tension through situated progressive disclosure. \name organizes each scenario into a decision tree of investigative choices. Players select actions (e.g., check who has access to these data or review the sign-up flow) that progressively reveal system behaviors. This requires players to determine which details warrant scrutiny, exercising \textit{selective attention}, while limiting information exposure to the chosen investigative path. To further reduce the cognitive overhead of tracking evidence across steps, \name automatically accumulates evidence as players progress (Figure~\ref{fig:screenshots}b).

\sssec{Templated Reasoning.} Two challenges undermine developers' ability to reason about privacy violations. First, many developers lack knowledge of what constitutes a privacy violation in practice~\cite{alhazmi2021m,alomar2022developers}. Second, even when developers sense that something is problematic, they struggle to articulate which specific decisions constitute the violation and what evidence supports the claim~\cite{horstmann2024those}. 

We address both challenges through violation report templates. Players can report a suspected violation at any point. They start by selecting from a menu of privacy violations. Each type is named and defined, making violation categories legible without requiring prior knowledge. They then construct a justification by placing evidence into predefined slots that operationalize the violation as a simple schema. For example, a \textit{Misrepresentation of Practices} violation~\cite{unfairftc} requires one evidence piece for the \textit{Claim} (the consumer-facing statement, promise, or impression created by the company) and one for the \textit{Actual Practice} (what the company does that conflicts with the claim) (Figure~\ref{fig:screenshots}d). The template directs players toward the specific evidence a valid argument requires. When a required slot cannot be filled, it prompts players to reconsider rather than over-flag.

\sssec{Feedback on Report Submissions.} Mistakes in violation reports reveal where players' reasoning breaks down, making each mistake a concrete opportunity for learning~\cite{epstein2002immediate}. With generic correct/incorrect feedback, however, players have no basis for directed revision and may resort to random adjustment. Instead, \name classifies each submission into one of four outcomes -- \textit{exact match}, \textit{near match}, \textit{reasoning misalignment}, and \textit{no match} -- and delivers targeted guidance immediately after each attempt. Each outcome directs players to the specific element that needs revisiting. For example, a near-match response affirms the violation type while redirecting the player to reconsider the one evidence slot that is incorrect. For a correct submission, feedback confirms the reasoning and connects the identified practice to its legal basis, reinforcing why the practice constitutes a violation. Table~\ref{tab:feedback-mechanism} details the definition and feedback strategy for each of the four outcomes.

\begin{table*}[h]
\centering
\small
\renewcommand{\arraystretch}{0.9}
\newcolumntype{P}[1]{>{\RaggedRight\arraybackslash\hspace{0pt}}p{#1}}
\caption{Feedback messages shown after each submitted violation report. The game classifies a player’s submission into one of four outcomes and presents targeted guidance with increasing specificity.}
\begin{tabularx}{\linewidth}{@{} P{0.08\linewidth} P{0.23\linewidth} P{0.21\linewidth} P{0.4\linewidth} @{}}
\toprule
\textbf{Outcome} & \textbf{Definition} & \textbf{Feedback Strategy} & \textbf{Feedback Example} \\
\midrule
Exact match & 
The submitted violation type and all evidence slot assignments match an unresolved ground-truth violation. & 
Summarize the violation, then identify the regulatory requirement it violates. & 
\textbf{Congratulations!} Premom promised not to share health data without consent while transmitting plain-text events with descriptive titles like ``Log period-save'' to Google and AppsFlyer. This misrepresentation of practices is a deceptive practice prohibited under FTC Act §5(a).
\\
\midrule
Near-match & 
The submitted violation type matches an unresolved ground-truth violation, and exactly one evidence slot contains an incorrect card. & 
Acknowledge the near-match and guide players to revisit the missed field without revealing the answer. & 
\textbf{So close!} You've correctly identified that there's a misrepresentation of practice here. Rethink your Claim: what was promised about data sharing in the company's consumer-facing statements.
\\
\midrule
Reasoning misalignment & 
The set of submitted evidence cards matches an unresolved ground-truth violation, but the submission does not qualify as an exact or near match. & 
Affirm the evidence set and prompt players to remap each clue to its role. & 
\textbf{You’ve got the pieces.} You’ve collected the right clues, but some roles don’t fit. Rethink about how the evidences connect to each other.
\\
\midrule
No match & 
The submission does not qualify as any of the above categories against any unresolved ground-truth violation. & 
Direct the player toward an unidentified violation. & 
\textbf{Not a match yet.} There's a misrepresentation of practices you haven't captured. Review the data sharing to see whether any consumer-facing promises conflict with what actually happened.
\\
\bottomrule
\end{tabularx}
\label{tab:feedback-mechanism}
\end{table*}

\sssec{FTC Counts as Ground Truth.} Privacy sensitivity calibration requires a credible reference point against which developers can evaluate whether their threshold for concern is appropriate. Expert opinion is one candidate, but it is difficult to scale. We use the violation counts in FTC enforcement documents as the ground truth to calibrate players' privacy sensitivity. Each case is based on a real incident in which regulators determined that a company's data practices caused—or posed a serious risk of—harm. This anchors the game's ``correct'' answers in concrete enforcement outcomes, so players learn which practices constitute actionable privacy violations rather than debating hypothetical edge cases. Section~\ref{sec:legaldocs} describes how we translate the allegations and evidence in FTC documents into game scenarios and map them to violation types and required evidence.

%% file: parts/p4-harness-legal-documents.tex
\section{Building from FTC Enforcements}\label{sec:legaldocs}

This section describes how we derive game scenarios, evidence, and ground truth from public U.S. FTC enforcement documents.

\subsection{Background: U.S. FTC Enforcements}

In the United States, the FTC is the primary federal agency responsible for enforcing consumer protection laws, including those related to privacy and data security. The FTC can pursue enforcement when companies engage in unfair or deceptive privacy practices—for example, misrepresenting how they collect, use, or share personal data. 
Many privacy cases end in public settlement orders that legally bind the company to stop the challenged practices and implement corrective measures. To date, the FTC’s website lists 328 cases tagged under privacy and security~\cite{CasesbyT26:online}.

These FTC enforcement documents follow a consistent, highly structured format that explicitly sets out both the challenged data practices and the legal reasoning in each case (Table~\ref{tab:ftc-complaint-sections}). They typically begin with a \textit{Summary of the Case} and a \textit{Defendant Description}, which provide high-level context about the company and the product or service at issue. The core \textit{Allegations} section then describes the disputed practices in concrete, specific terms—often naming what data was collected or shared, through which technical mechanisms (e.g., SDKs, logging, transmissions), what the company told users, and what harms or risks could result. These descriptions are generally written in plain, accessible language rather than technical or legal jargon, making them easy to understand for non-lawyers. The allegations are mapped to one or more \textit{Violation Counts}, which state how the practices constitute unfair or deceptive acts under particular legal provisions, and the documents conclude with a \textit{Prayer for Relief} specifying the remedies sought.

\begin{table*}[h]
\small
\renewcommand{\arraystretch}{0.9}
\centering
\caption{Typical content structure of FTC enforcement documents. Examples are drawn from~\cite{EasyHeal44:online}.}
\begin{tabular}{p{0.09\linewidth} p{0.18\linewidth} p{0.65\linewidth}}
    \toprule
    \textbf{Section} & \textbf{Definition} & \textbf{Example Content} \\
    \toprule
    Summary of the Case &
    A brief overview about the case &
    ``Between 2017 and 2020, Defendant repeatedly and falsely promised Premom users in their privacy policies that Defendant: (a) would not share health information with third parties without users’ knowledge or consent; (b) to the extent Defendant collected and shared any information, it was non-identifiable data; and (c) the data was used only for Defendant’s own analytics or advertising.'' \\
    \midrule

    Defendant Description &
    The identification about the company &
    ``Defendant Easy Healthcare Corporation (``Easy Healthcare'') is an Illinois corporation with its principal office or place of business at 360 Shore Dr. Unit B, Burr Ridge, IL 60527.'' \\
    \midrule

    Allegations &
    A detailed report of the defendant's actions and how those actions can harm consumers &
    ``Defendant failed to take reasonable measures to assess and address privacy risks to user information while creating and maintaining Premom. For example: (a) Defendant failed to adequately assess the privacy risks of third-party SDKs prior to incorporating those SDKs into Premom; \ldots''

    ``As a result of these practices, any third party who acquired this data, including foreign governments or bad actors, could decrypt and access Premom users’ sensitive data, including precise geolocation information and non-resettable identifiers \ldots'' \\
    \midrule

    Violations Count &
    A narrative stating how the defendant broke the law &
    ``Defendant’s failure to disclose the material information described in Paragraph 81 constitutes a deceptive act or practice in violation of Section 5(a) of the FTC Act, 15 U.S.C. \S\ 45(a).'' \\
    \midrule

    Prayer for Relief &
    The plaintiff requests the Court for remedies &
    ``Award Plaintiff monetary civil penalties from Defendant for each violation of the Health Breach Notification Rule alleged in this Complaint.'' \\
    \bottomrule
\end{tabular}

\label{tab:ftc-complaint-sections}
\end{table*}

\subsection{Templates for Privacy Violation Reasoning}

We iteratively developed a set of violation templates to represent privacy violations by analyzing 40 of the most recent privacy enforcement cases published on the FTC website. The goal of these violation templates is twofold: (1) to scaffold the process of searching for potential violations in complex, real-world materials by making salient what to look for, and (2) to support privacy reasoning by structuring how evidence is linked to the underlying privacy principle or regulatory requirement.

To construct the violation templates, we conducted a series of collaborative card-sorting sessions~\cite{righi2013card, wood2008card}. Two participants participated across 3 iterative sessions. Both participants had hands-on experience reading FTC enforcement documents through prior projects, giving them familiarity with recurring regulatory frameworks; one brought deeper regulatory reading experience while the other contributed a developer's perspective aligned with the game's target audience. In each iteration, participants first independently proposed templates based on the enforcement documents, then engaged in group discussion to reconcile differences and converge on a shared set. The converged set from each session served as the starting point for the subsequent iteration.

Through this iterative process, we observed that all FTC allegations are explicitly grounded in specific regulatory requirements, such as FTC Act Section 5(a), COPPA, or ROSCA. Each requirement translates an established privacy principle, such as informed consent, notice, into an enforceable legal obligation. Individual enforcement cases often contain multiple distinct violations. For each violation, the complaint typically specifies: (1) the data practice at issue, (2) the consequence of the practice, (3) the violation claims of certain regulations or privacy principles, (4) the specific evidence supporting the claim, and (5) how the FTC requires the company to remediate the violation. However, remediation requirements and consequences vary widely across cases and are highly context-dependent, making them difficult to encode in a compact, generalizable representation.

Based on these observations, we focused the violation templates on the components that were both consistently present and central for reasoning: why a practice is problematic and the evidence supporting that determination. We found that many allegations can be encoded using structured templates derived from the underlying regulation cited in the complaint. For example, in the Alexa enforcement Count IV~\cite{US_v_Amazon_Alexa_2023}, the FTC alleges violations of the Children’s Online Privacy Protection Act (COPPA) related to COPPA § 312.10 (data retention and deletion requirements), which indicates that operators may retain children’s personal information only for as long as reasonably necessary to fulfill the stated purpose. Abstracting from this requirement, the violation can be represented as excessive retention of children’s data, which requires two key evidentiary elements: (1) evidence that personal information was collected from children and (2) evidence describing the duration or conditions of retention.

We operationalized this abstraction process in three steps. First, for each regulation cited in an FTC complaint, we reviewed the regulatory text (including relevant definitions and related provisions) and derived an abstract violation template that captures the core prohibited pattern (i.e., what must be shown for the violation to hold). Second, we validated each abstraction by instantiating it on the corresponding enforcement “count”: we filled the template with case-specific, contextualized descriptions drawn directly from the complaint (e.g., what data were involved, who was affected, what the company claimed or did, and what evidence the FTC cited). 
Third, we expanded our analysis across additional enforcement cases and iteratively consolidated overlapping templates. Through this process, we found that most privacy allegations in the dataset repeatedly relied on a small set of recurring legal grounds, ultimately converging on nine commonly cited privacy regulations. Appendix Table~\ref{tab:principle-schema} summarizes the resulting violation templates.

\begin{figure*}[h!]
    \centering
    \includegraphics[width=1\linewidth]{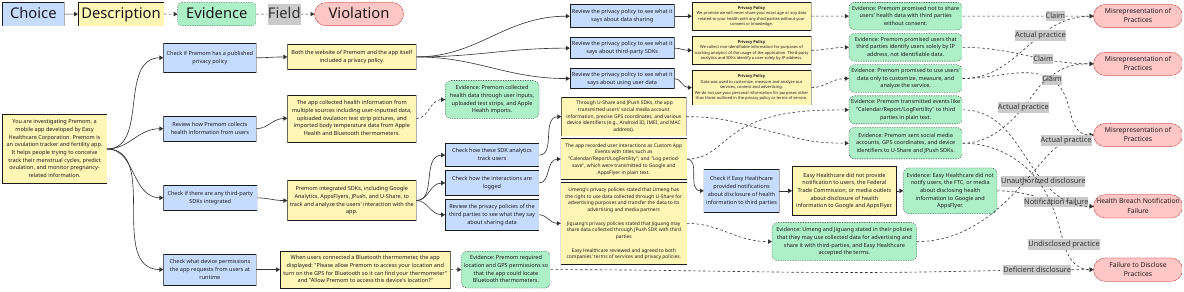}
    \caption{Progressive tree structure for in-game investigation. Solid lines indicate how descriptions are connected by choices, which players select to explore additional information. Dashed lines link descriptions to their associated evidence, which players automatically receive upon reaching the corresponding description. Connections between evidence and violations constitute the correct solution to the reasoning task and must be inferred by players during the gameplay session.}
    \label{fig:tree-structure}
\end{figure*}

\subsection{Game Scenario for Progressive Search}

To support progressive exploration, we further transformed FTC settlement documents into a tree-structured representation composed of descriptions, choices, evidence, and violations (Figure~\ref{fig:tree-structure}).

\begin{figure}[h]
    \centering
    \includegraphics[width=1\linewidth]{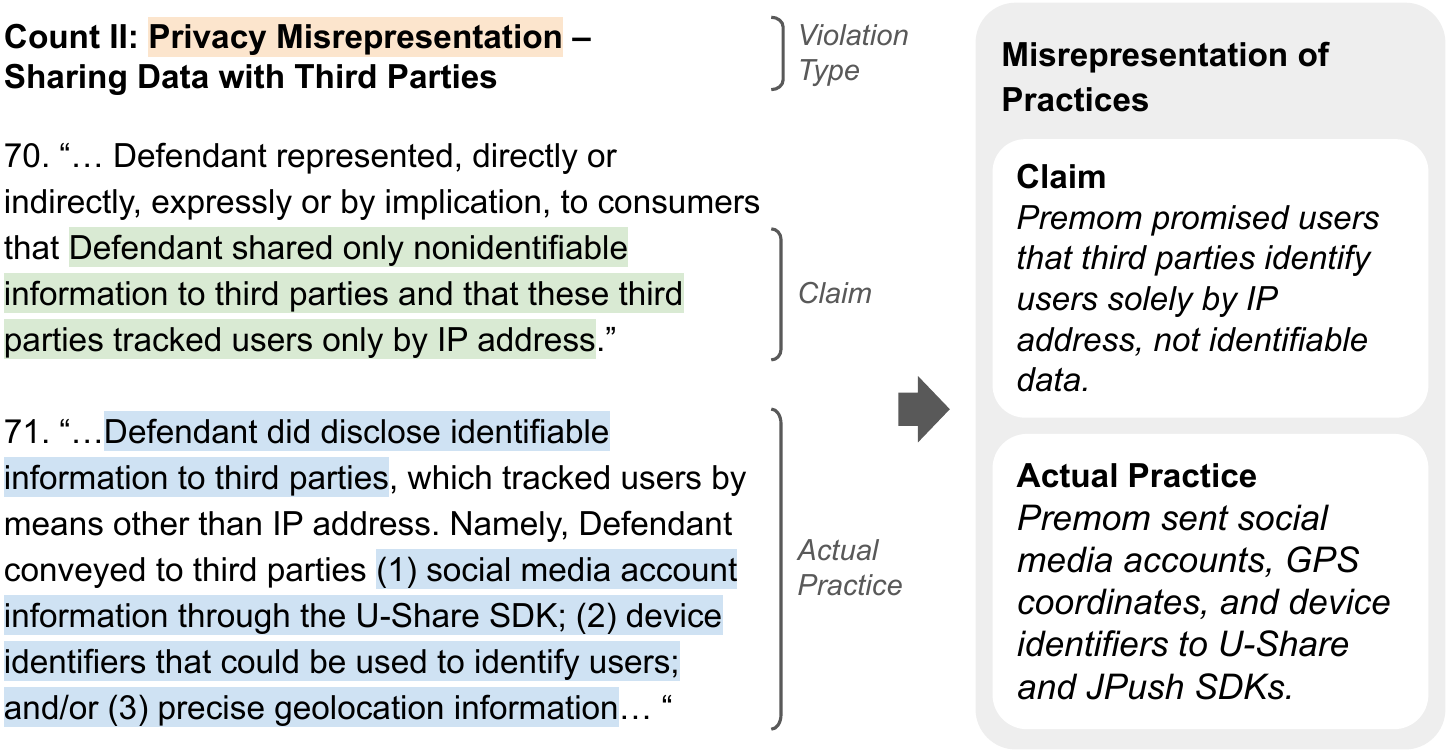}
    \caption{We used ChatGPT to paraphrase text from enforcement documents into in-game narratives, followed by manual refinement.}
    \label{fig:extraction}
\end{figure}

At the core of the tree are descriptions, which summarize factual information from the complaint and explain what happened in the case. These descriptions are not verbatim excerpts from enforcement documents; instead, they are synthesized from information that is often dispersed across multiple sections of the allegations (Figure~\ref{fig:extraction}). 
Attached to each description are evidence cards, which provide an even more concise summary of specific facts relevant to potential violations. We intentionally kept evidence cards short so that players could easily revisit and compare them when matching evidence to violation templates. While descriptions may include contextual or narrative information, evidence cards focus strictly on facts that directly support or refute a violation and exclude background material.

Choices connect descriptions into a decision tree and represent the player’s actions as an investigator. Rather than presenting all information at once, we decomposed complex cases into smaller, thematically grouped descriptions (e.g., what the company disclosed in its privacy policy, how data were logged via SDKs, what permissions were requested at runtime). Players select choices corresponding to investigative activities—such as reviewing privacy policies, monitoring network traffic, reverse-engineering third-party SDK usage, or inspecting permission requests—which determine which description they encounter next (Figure~\ref{fig:screenshots}a). We developed these choices by manually clustering related evidence and mapping them to plausible investigative steps.
Finally, violations correspond to the “counts” in the enforcement. Each violation is associated with a violation type (i.e., the regulatory requirement at issue) and is supported by multiple pieces of evidence.

All descriptions and evidence were generated by summarizing and rewriting the relevant portions of the complaint using ChatGPT, followed by manual review and refinement to ensure clarity and comprehensibility for players.

%% file: parts/p5-design-iteration.tex
\section{Pilot Tests, Iterations, and Implementation}

We used an interactive prototyping process~\cite{fitton2005rapid} to develop the system: we built early versions of the game as a web-based game, tested them with internal users, collected structured feedback, and iteratively refined the design.
These pilots led to several improvements. First, we made feedback more direct, particularly for ``near-match'' responses, by explicitly identifying the missing element in player's reasoning. Second, we added a lightweight glossary to support participants with diverse levels of technical maturity (e.g., brief explanations of terms such as ``SDK''). Third, to maintain engagement and guide attention, we implemented a character-by-character streaming display and revealed response options only after the relevant text was fully shown. We also added animations when evidence is collected and incorporated a submission history to help players track prior attempts. Finally, we retained real company names because pilots indicated that realism and surprise increased engagement, and we addressed additional usability issues identified during testing.

%% file: parts/p6-evaluation.tex
\section{Evaluation}

We conducted a between-subjects user study to evaluate how \name compares with a reading-based learning approach.

\subsection{Study Design}

\begin{table*}[h!]
\caption{We used three case tracks, each with two FTC cases: one for pretest/learning and one for posttest. We paired cases that share violation types but differ in context. Each bullet (\textbullet) represents one violation instance.}
\label{tab:cases}
\newcommand{\D}{\textbullet}
\newcommand{\nd}{-}
\resizebox{\textwidth}{!}{%
\setlength{\tabcolsep}{4pt}
\begin{tabular}{lcccccc}
\toprule
\textbf{Case Track} & \multicolumn{2}{c}{\textbf{Health Data Practices}}
& \multicolumn{2}{c}{\textbf{Children's Data Practices}}
& \multicolumn{2}{c}{\textbf{Dark Patterns}} \\
\cmidrule(lr){2-3}\cmidrule(lr){4-5}\cmidrule(lr){6-7}
\textbf{Phase} & Pretest/Learning & Posttest
& Pretest/Learning & Posttest
& Pretest/Learning & Posttest \\
\cmidrule(lr){2-3}\cmidrule(lr){4-5}\cmidrule(lr){6-7}
\textbf{Enforcement Case}
& \makecell{\textit{Easy Healthcare}\\\cite{US_v_EasyHealthcare_2023}}
& \makecell{\textit{GoodRx}\\\cite{US_v_GoodRx_2023}}
& \makecell{\textit{Alexa}\\\cite{US_v_Amazon_Alexa_2023}}
& \makecell{\textit{Musical.ly}\\\cite{US_v_Musically_2019}}
& \makecell{\textit{Amazon}\\\cite{US_v_Amazon_Rosca_2023}}
& \makecell{\textit{Age of Learning}\\\cite{US_v_Age_of_Learning_2020}} \\
\midrule
Misrepresentation of Practices                   & \D\,\D\,\D & \D\,\D\,\D\,\D\,\D & \D\,\D & \nd & \nd & \D \\
Failure to Disclose Practices                    & \D & \nd & \nd & \D & \D\,\D\,\D & \D \\
Health Breach Notification Failure               & \D & \D & \nd & \nd & \nd & \nd \\
Excessive Retention of Children's Data           & \nd & \nd & \D & \D & \nd & \nd \\
Failure of Parental Control over Children's Data & \nd & \nd & \D & \D & \nd & \nd \\
Failure of Parental Notice about Children's Data & \nd & \nd & \D & \D & \nd & \nd \\
Failure to Obtain Verifiable Parental Consent    & \nd & \nd & \nd & \D & \nd & \nd \\
Failure to Clearly Disclose Transaction          & \nd & \nd & \nd & \nd & \D & \D \\
Failure to Obtain Express Informed Consent       & \nd & \nd & \nd & \nd & \D & \D \\
Failure to Provide Simple Cancellation Mechanism & \nd & \nd & \nd & \nd & \D & \D \\
\bottomrule
\end{tabular}%
}
\end{table*}

\sssec{Baseline selection.} In the reading-based condition, participants learned from each FTC enforcement case by reading both the official press release and the complaint.

\revision{We selected this baseline for two reasons. First, it reflects the dominant way developers develop privacy awareness: by accessing online resources such as news articles, technical blogs, and coverage of real incidents~\cite{tahaei2021privacy, prybylo2024evaluating}, rather than through formal training. Second, we grounded both conditions in the same source documents to keep content exposure consistent across conditions. Alternative interventions, such as concept-based quizzes~\cite{android_privacy_security_quiz} or role-playing simulations~\cite{shilton2020role}, would introduce confounds in the content. As a result, we could not cleanly isolate the effect of the game’s interactive mechanics.}

In our baseline, the FTC press release serves as a news-style entry point, while the publicly accessible complaint provides detailed data practices and alleged violations. We instruct participants to read these materials with the goal of understanding what privacy violations may have occurred in the case. Complaints can be lengthy, spanning tens of pages. We do not ask participants to read every line, but instead skim and focus on sections they consider most relevant for identifying and understanding potential violations. We enforce the same time limit on the learning phase in both conditions.

\sssec{Case assignment.} We assigned participants to one of three \emph{case tracks} (Table~\ref{tab:cases}) using stratified randomization~\cite{suresh2011overview}, ensuring that each track included equal numbers of participants in the game and reading conditions. Each case track consists of two FTC enforcement cases. The first case serves as both the pretest and learning case, and the second case serves as the posttest case. We selected the two cases with overlapping violation types but differing contexts to test whether participants transfer their reasoning about those principles to new scenarios. We tried our best to find cases with identical sets of violations, but it is rare for real-world enforcement actions to fully align.

\subsection{Study Procedure}
Before starting the study, participants reviewed and signed a consent form. The form explained the study's purpose, procedures, and data collection practices. It also noted that participation was voluntary and that participants could withdraw at any time without penalty.

\begin{figure}[h]
    \centering
    \includegraphics[width=1\linewidth]{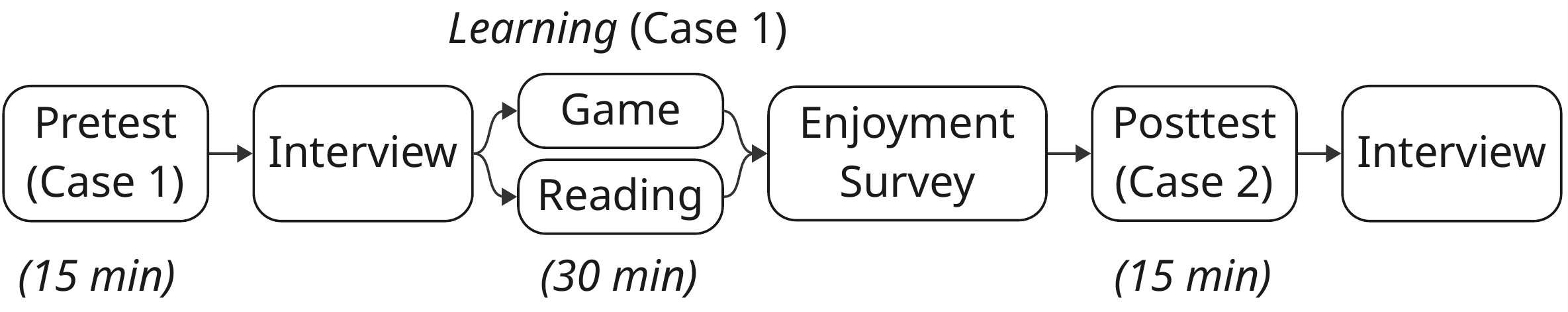}
    \caption{Study procedure.}
    \label{fig:placeholder}
\end{figure}

\sssec{Pretest \& Interview.} We asked participants to read a narrative of the first case in their case track and list all violations they believed applied, each with 1--2 sentences of reasoning. We derived the narrative from the complaint, rewriting it in a neutral tone that avoided explicit statements about negative consequences. The narrative summarized the case background and key design decisions about privacy policies, system behavior, and user interface design. The legal counts that participants were asked to identify were omitted. Appendix~\ref{app:narratives} presents the narrative we used for the \textit{Health Data Practices} track as an example.
To clarify what we meant by violations, we provided a list of violation types and their definitions, without the corresponding fields (Appendix~\ref{app:violation-type-list}). The list presented all violation types that appeared in all six cases. Each participant had 15 minutes to complete the pretest. We did not provide feedback on correctness after the pretest.

After participants submitted their responses, we conducted a brief interview ($\approx$ 5 minutes) about how they approached the task. We asked them to elaborate on their strategy for identifying violations and to highlight any violation types they felt particularly confident or uncertain about in this case.

\sssec{Learning \& Survey.} Participants then studied the same case using their assigned learning method. In the game condition, participants played \name built around the case with a goal of finding as many violations as possible while minimizing the number of attempts. In the reading condition, participants to read the FTC press release and then skim the complaint, focusing on sections they found relevant to potential violations. The learning phase ended when participants chose to move on or when they reached the 30-minute time limit. 

After completing the learning, participants filled out an enjoyment survey. We measured learner enjoyment using the Interest/Enjoyment subscale of the Intrinsic Motivation Inventory (IMI)~\cite{ryan1982control}, rated on a 7-point Likert scale, which captures whether learners found the activity enjoyable and felt motivated to learn.

\sssec{Posttest \& Interview.} \revision{After learning, participants completed a posttest on the second case. The posttest followed the same format and 15-minute limit as the pretest, while participants had no access to prior materials. We then conducted the same brief interview as after the pretest.}

\sssec{Format \& Data Collection.} We ran all study sessions in person and asked participants to join a Zoom meeting and share their screens. We recorded screen video and audio and logged responses to the pretest, posttest, and surveys, along with time spent in each study section. In the game condition, we also recorded in-game performance as the number of correctly reported violations.

\subsection{Participants}

\revision{We recruited 18 participants per condition (36 total), following prior formative evaluations of educational interventions~\cite{sheng2007anti, wen2019hack} that prioritize an initial empirical signal over a definitive efficacy estimate. We recruited student developers with hands-on programming experience as our target participants. We advertised the study via department Slack channels and campus flyers.} To avoid attracting mostly participants who already care about privacy, we advertised the study as ``reviewing software design'' rather than explicitly referencing privacy. The sign-up survey collected respondents’ contact information and background information, including demographics and programming experience. We screened respondents and recruited only currently enrolled students in computer science-related majors who had at least two years of programming experience.

Our final participant pool included 20 undergraduates and 16 graduate students, predominantly CS majors (63.9\%), male-identifying (61.1\%), and aged 18--24 (86.1\%). Table~\ref{tab:demographics-breakdown} presents the demographics breakdown by condition.

% \begin{table}[h!]
% \centering
% \footnotesize
% \renewcommand{\arraystretch}{0.9}
% \caption{Participant demographics.}
% \begin{tabular}{llll}
% \toprule
% & \textbf{Category} & \textbf{Game} & \textbf{Reading} \\
% \midrule

% \multirow[t]{4}{*}{Gender}
% & Male & 10 (55.6\%) & 12 (66.7\%) \\
% & Female & 6 (33.3\%) & 5 (27.8\%) \\
% & Non-binary & 1 (5.6\%) & -- \\
% & Prefer not to say & 1 (5.6\%) & 1 (5.6\%) \\
% \midrule

% \multirow[t]{2}{*}{Age}
% & 18--24 & 16 (88.9\%) & 15 (83.3\%) \\
% & 25--34 & 2 (11.1\%) & 3 (16.7\%) \\
% \midrule

% \multirow[t]{3}{*}{Education}
% & Undergraduate & 10 (55.6\%) & 10 (55.6\%) \\
% & Graduate & 8 (44.4\%) & 8 (44.4\%) \\
% \midrule

% \multirow[t]{5}{*}{Major}
% & Computer Science & 10 (55.6\%) & 13 (72.2\%) \\
% & Data Science & 4 (22.2\%) & 2 (11.1\%) \\
% & Elec. \& Comp. Eng. & 2 (11.1\%) & 1 (5.6\%) \\
% & Bioinformatics & 1 (5.6\%) & 1 (5.6\%) \\
% & Mathematics & 1 (5.6\%) & -- \\
% & Cognitive Science & -- & 1 (5.6\%) \\
% \midrule

% \multirow[t]{4}{*}{Prog. Exp.}
% & More than 5 years & 4 (22.2\%) & 4 (22.2\%) \\
% & 4--5 years & 8 (44.4\%) & 4 (22.2\%) \\
% & 3--4 years & 4 (22.2\%) & 5 (27.8\%) \\
% & 2--3 years & 2 (11.1\%) & 5 (27.8\%) \\
% \bottomrule
% \end{tabular}
% \label{tab:demographics-breakdown}
% \end{table}

\subsection{Analysis Method} \label{evaluation-analysis}

\sssec{Evaluation Metrics.} We evaluated participants’ open-ended pretest and posttest responses by comparing the reported violation instances against those alleged in the FTC complaints, which we treated as ground truth. For each ground-truth instance, we specified its violation type and key characterizing elements (Appendix~\ref{app:narratives} shows an example). We then coded each reported violation as a binary classification: a \textit{true positive (TP)} if it clearly referred to the same underlying issue as a ground-truth instance, or a \textit{false positive (FP)} otherwise.

Two researchers with one and two years of experience in privacy research, respectively, served as coders and co-developed the codebook. In a pilot phase, both independently coded the same subset of responses, then met to review disagreements and refine the codebook until a shared interpretation was reached. For the final coding, both coders independently coded each response using the finalized codebook, yielding two labels per reported-violation instance with no missing labels. We computed Krippendorff’s $\alpha$ on the per-instance labels~\cite{hayes2007answering} and obtained $\alpha = 0.74$, indicating moderate agreement~\cite{krippendorff2018content}. Coders resolved remaining disagreements through discussion and all analyses used the adjudicated codes.

After coding, we computed \textit{false negatives (FN)} as the number of ground-truth violation instances minus the number of TP. We then derived recall and precision from the TP, FP, and FN counts.

\sssec{Quantitative Data Analysis.} To test whether training improved participants' ability to identify privacy violations, we fit two linear mixed-effects regression model with \textit{recall} and \textit{precision} as dependent variables. We chose mixed-effects regression as each participant contributed two observations (pretest and one posttest), introducing within-person correlation that simple regression cannot capture. A participant-level random intercept accounts for such dependency. Fixed effects included condition (pretest, post-reading, post-game) and case track (Health, Children, Dark Patterns), with categorical predictors dummy-coded using $N-1$ indicators. We verified residual normality via Q-Q plots, which showed approximate normality across all outcomes (Appendix~\ref{app:qq-plots}).

We prioritized recall because it reflects participants' coverage of known violations, determining what gets escalated to privacy experts. We also analyzed precision to ensure that flags remained targeted rather than applied to non-issues.

We computed each participant’s enjoyment by reverse-coding the two negatively worded items (subtracting each response from 8) and averaging all six items. We used Mann–Whitney U tests to test for between-condition differences as Shapiro-Wilk tests showed non-normal distributions ($p < 0.05$). We report the median, mean, and standard deviation for each condition.

\sssec{Qualitative Data Analysis.} We analyzed interview responses using an open coding approach, organizing participants' reported strategies and self-described learning into categories. We transcribed recordings with Zoom’s auto-transcription, then one author verified and corrected each transcript. The same two coders from the quantitative analysis independently coded the 10 of 36 transcript pairs to develop an initial codebook, then met to incorporate codes and reconcile differences. They then divided the remaining 26 transcript pairs, each coded by one coder, and resolved disagreements through discussion. Because our goal was to reach a shared interpretive framework rather than measure coder independence, we did not compute IRR~\cite{mcdonald2019reliability}. The jointly coded subset served to surface disagreements and stabilize the codebook. These findings are exploratory and intended to complement rather than independently validate our quantitative results. We present the complete codebook in Appendix~\ref{app:qualitative-codebook}.

\subsection{Quantitative Results}

\begin{table*}[h]
\centering
\footnotesize
\renewcommand{\arraystretch}{1.2}
\caption{Linear mixed-effects model coefficients, p-values, and 95\% CIs. Positive coefficients indicate higher metric scores. Post-game scores increase significantly (+0.242 recall, +0.127 precision, +0.219 reasoning completeness), whereas post-reading changes are not significant.}
\setlength{\tabcolsep}{4.5pt}
\begin{tabular}{llllllllll}
\toprule
Variable & \multicolumn{3}{c}{Recall} & \multicolumn{3}{c}{Precision} & \multicolumn{3}{c}{Reasoning completeness} \\
\cmidrule(lr){2-4} \cmidrule(lr){5-7} \cmidrule(lr){8-10}
 & Coeff. & $p$-value & 95\% CI & Coeff. & $p$-value & 95\% CI & Coeff. & $p$-value & 95\% CI \\
\midrule
\textbf{Condition}: \textbf{Post-game (vs. Pre)}    & \textbf{0.242} & \textbf{0.000$^{***}$} & \textbf{[0.125, 0.359]} & \textbf{0.127} & \textbf{0.033$^{**}$}  & \textbf{[0.010, 0.245]} & \textbf{0.219} & \textbf{0.001$^{***}$} & \textbf{[0.085, 0.352]} \\
\textbf{Condition}: Post-reading (vs. Pre) & 0.099 & 0.099$^{*}$   & [-0.019, 0.216] & 0.005 & 0.930        & [-0.112, 0.123] & 0.071 & 0.299         & [-0.063, 0.204] \\
\textbf{Case}: Children (vs. Health)                  & 0.120 & 0.070$^{*}$   & [-0.010, 0.249] & -0.111 & 0.064$^{*}$ & [-0.228, 0.006] & -0.062 & 0.480        & [-0.236, 0.111] \\
\textbf{Case}: Dark Patterns (vs. Health)                  & 0.045 & 0.499         & [-0.085, 0.174] & 0.079  & 0.186       & [-0.038, 0.196] & -0.115 & 0.194        & [-0.289, 0.059] \\
\textbf{Intercept}                          & 0.512 & 0.000$^{***}$  & [0.409, 0.615]  & 0.776 & 0.000$^{***}$ & [0.681, 0.872] & 0.631 & 0.000$^{***}$ & [0.497, 0.764] \\
\textbf{Random effect}                      & 0.005 &               &               & 0.000 &             &               & 0.022 &              &               \\
\midrule
Marginal $R^2$ / Conditional $R^2$  & \multicolumn{3}{r}{0.223 / 0.314} & \multicolumn{3}{r}{0.174 / 0.174} & \multicolumn{3}{r}{0.218 / 0.456} \\
\bottomrule
\\
\multicolumn{10}{l}{\small Note: * $p < 0.1$, ** $p < 0.05$, *** $p < 0.01$} \\
\end{tabular}
\label{tab:lme-recall-precision-reasoning}
\end{table*}

\begin{figure}[h]
    \centering
    \includegraphics[width=1\linewidth]{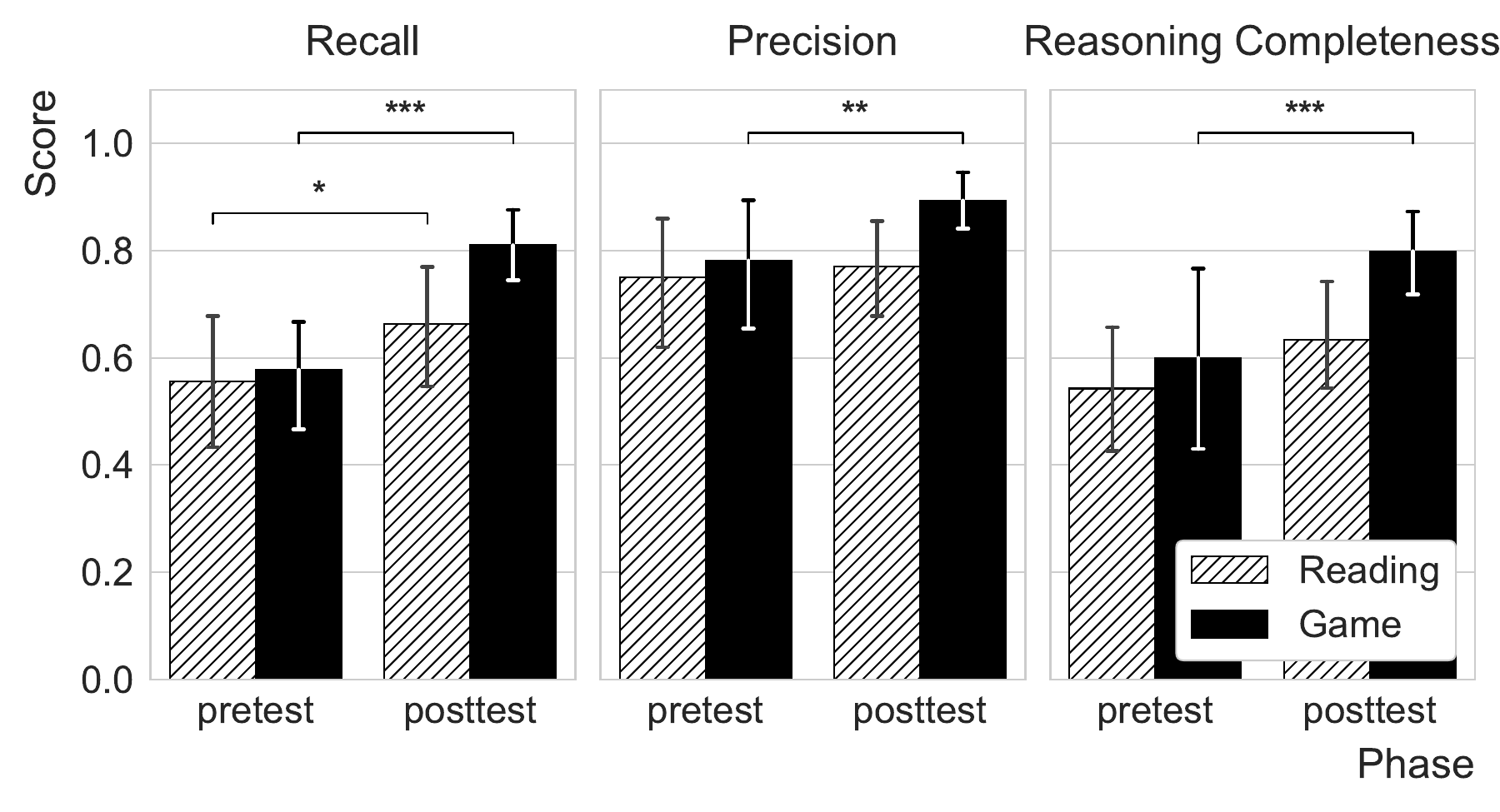}
    \caption{From pretest to posttest, the game group identified more true violations (higher recall), reported a higher share of ground-truth violations (higher precision), and provided more complete justifications. The reading group showed a marginal improvement in recall ($p < .10$) but no significant gains in precision or reasoning completeness. Error bars indicate 95\% confidence intervals.}
    \label{fig:learning-effectiveness-bar-charts}
\end{figure}

\sssec{\name produced a significant improvement in recall, whereas reading did not.} This pattern is unlikely due to baseline differences as the two groups did not differ in pretest recall (Mann--Whitney U, $p=0.667$). The linear mixed-effects model (Table~\ref{tab:lme-recall-precision-reasoning}) shows that recall increased significantly from pretest to post-game ($\beta=0.242$, 95\% CI \mbox{[0.125, 0.359]}, $p<0.001$), whereas the post-reading increase was smaller and did not reach the $\alpha=0.05$ threshold ($\beta=0.099$, 95\% CI \mbox{[-0.019, 0.216]}, $p=0.099$). \revision{The 95\% confidence interval for the post-game recall effect excludes zero at both bounds, indicating that the observed improvement is unlikely to reflect sampling variability alone.} The model’s marginal $R^2=0.223$ indicates that the fixed effects account for 22.3\% of the variance in recall, while the conditional $R^2=0.314$ shows that including participant-level random effects explains 31.4\%. Given the inherent noise in human behavior, $R^2$ values above 0.1 are often considered acceptable when key predictors are statistically significant~\cite{ozili2023acceptable}. Figure~\ref{fig:learning-effectiveness-bar-charts} visualizes these trends.

\sssec{Higher recall after playing \name did not come at the cost of over-flagging.} Recall can be inflated by flagging more issues overall, which increases hits but also false positives. We used precision to check for over-flagging as precision penalizes false positives relative to true positives. We found no pretest difference between groups (Mann--Whitney U, $p=0.633$). The mixed-effects model (Table~\ref{tab:lme-recall-precision-reasoning}) then showed that precision increased after gameplay relative to pretest ($\beta=0.127$, 95\% CI \mbox{[0.010, 0.245]}, $p=0.033 < 0.05$), while reading produced no significant improvement ($\beta=0.005$, $p=0.930$). 

\sssec{Participants in the game group increased the completeness of their justifications.} We defined reasoning completeness as participants' ability to justify identified violations using correct violation types and key evidence pieces. We operationalized it by splitting true positives into \textit{fully justified} and \textit{partially justified}: \textit{fully justified} use the correct violation type and include all key evidence, while \textit{partially justified} flag the same issue but use the wrong type label or omit one key element. We computed reasoning completeness as the ratio of \textit{true positives} that qualified as \textit{fully justified}. We used consensus coding to reach collective agreement rather than estimate inter-rater reliability.

Before training, the two groups showed comparable completeness (Mann--Whitney U, $p=0.502$). Using the same linear mixed-effects model as our recall and precision analysis, we found that \name significantly increased reasoning completeness ($\beta=0.219$, 95\% CI \mbox{[0.085, 0.352]}, $p=0.001$), whereas reading did not ($\beta=0.071$, $p=0.299$; Table~\ref{tab:lme-recall-precision-reasoning}). Overall, \name improved both participants' coverage of privacy violations and the completeness of their justifications.

\sssec{Enjoyment was comparable across conditions.} Participants reported moderately high enjoyment in both the game condition (median=5.55, mean=5.26, SD=1.34) and the reading condition (median=5.10, mean=4.82, SD=1.36), with no significant difference between groups (Mann-Whitney U, $p=0.288$). Within the game condition, participants who found all five violations tended to rate enjoyment somewhat higher (median=5.90, mean=5.52, SD=1.41) than those who did not (median=4.40, mean=4.60, SD=0.97), though this difference was not significant ($p=0.205$). Given \name shows players their progress (number of violations found and remaining), this pattern may reflect differences in perceived competence, which prior work associates with enjoyment in educational games~\cite{grasse2022using,touati2018leads}.

\subsection{Qualitative Results}

\revision{Participants' interview responses provide exploratory insights into how they found violations.} We focus on the strategies they used and how these strategies shifted after training. In the following results, we refer to participants in the game condition as G1--G18 and those in the reading condition as R1--R18.

\sssec{\revision{Participants reported ad hoc attention strategies when identifying violations.}} For example, 12 participants (G1, G7, G8, G9, G14, G17, G18, R4, R10, R12, R17, R18) mentioned ruling out violation types that appeared irrelevant to the narrative. Following the Alexa case, G7 said, \textit{``there’s a lot of violation about health data or transactions in the list, but I don’t think they are relevant to this case.''} Ten participants (G8, G10, G11, G12, R1, R2, R9, R10, R11, R18) leveraged prior experience to interpret what practices seemed plausible or typical in real products. In response to the Amazon case in the pretest, G11 reasoned: \textit{``I don’t normally see any free trials that way, and I would imagine that if that was allowed, then maybe companies would have done that more. So given that they don’t do it, it’s probably also not allowed, and it also seems to fit into the category.''} 
These ad hoc processes sometimes mislead them into a wrong judgment.
For instance, after the Easy Healthcare case in the pretest, G1 explained, \textit{``I removed health breach since it’s not clear whether this is a breach,''} even though the case involved a health breach and the failure to provide notification. 
Twelve participants (G1, G3, G4, G5, G6, G16, G18, R1, R4, R8, R11, R14) recalled difficulty distinguishing among closely related violation categories. For example, in the Alexa case, G5 recognized that the issue involved children’s data but mixed up whether the problem was about parental consent, parental notice, parental control, or data retention, noting: \textit{``It's clear that the case has to do with some violation related to parental control. But I'm not sure which specific violation type should I put.''}

\sssec{\revision{Participants reported increased confidence across conditions, with game participants additionally describing more targeted evidence-seeking strategies.}} Eleven participants in the game group (G1, G2, G3, G4, G5, G6, G7, G9, G11, G14, G15) and nine in the reading group (R1, R3, R4, R5, R6, R7, R8, R12, R17) explicitly said they felt more confident in their judgments. For example, after the posttest, R4 recalled, \textit{``I felt more sure that this leans towards misrepresentation.''} Participants also described correcting earlier misconceptions. For instance, G1, who failed to recognize a Health Breach Notification Failure in the pretest, said in the posttest interview, \textit{``I originally understood health information breach differently. But now I know as long as it is shared without consent, it constitutes a health information breach.''}

In addition, eight participants (G1, G2, G3, G5, G6, G7, G13, G15) said they became more attentive to specific data-practice cues after seeing contextual signals (e.g., child-directed services), whereas no reading-group participant mentioned this shift. For example, after playing \name, G5 noted increased attention to children-facing products: \textit{``for products targeting children under 13, I'm now paying more attention on whether they provide parents control to their children's data and if parents received notice on children's data collection and deletion.''} Beyond identifying issues, seven game participants (G1, G3, G5, G6, G11, G13, G14) also reported adopting more explicit justification strategies. G1 explained that they learned to anchor a violation in specific evidence, such as pairing what was claimed with what actually occurred: \textit{``Now I know that I should look for specific parts when reasoning why the violation occur. Like the misrepresentation, I now know that I should put both the claim and what was actually did, instead of just paying attention to what they did.''}

%% file: parts/p7-discussion.tex
\section{Discussion}

\subsection{Designing for Privacy Awareness, Not Minimum Compliance}

\revision{Although we grounded \name in enforcement documents, our goal is not to train developers to optimize for legal permissibility. We view these documents not as compliance benchmarks but as harm-grounded post-mortems: the practices described in FTC complaints are not borderline cases -- they are practices regulators pursued precisely because they were associated with clear harm or serious risk of harm to real people. A ``minimum-compliance'' mindset can encourage boundary-pushing and turn privacy decision-making into a threshold game of how far a practice can go without triggering enforcement. By framing these documented harms as the lesson rather than the legal threshold as the target, \name cultivates awareness that steers developers toward user-respecting data practices, while leaving formal compliance analysis to privacy and legal specialists.}

\subsection{Using Game Format for Privacy Awareness Training}

Educational games have faced skepticism on limited learning gains~\cite{sitzmann2011meta}. Such approaches bet on motivational change: making learning fun increases engagement, which in turn produces learning~\cite{passarelli2019educational}. However, for complex reasoning tasks like identifying problematic privacy designs, even motivated learners need an environment that sustains their engagement long enough to practice their skills. For these complex tasks, the value of the game format does not lie in entertainment, but in its capacity to hold learners in the material long enough for the reasoning to develop.

\subsection{Why Does \name Lead to Better Learning Outcome?}

Unlike reading FTC documents, which presents violations as conclusions, \name's mechanics make privacy reasoning explicit and correctable. The violation templates require players to slot evidence into predefined fields, transforming a privacy argument into something to be constructed — an instance of representational guidance~\cite{ainsworth2006deft} that directs learners' attention to what matters rather than leaving them to infer it from prose. The game's targeted feedback then completes this by directing players to the specific locus of their error rather than simply confirming or denying correctness. This instantiates a productive failure dynamic~\cite{kapur2016examining}: generating imperfect arguments first activates prior knowledge and discovers gaps, preparing learners to consolidate the correct framework when feedback arrives. 
This is distinct from trial-and-error, where players are simply random guessing to quickly reach the right answer.

\subsection{Alternative Game Mechanics}

During development, we identified several directions for extending the game. The game could support multiple difficulty levels by scaling the evidence pool size, increasing the selective-attention burden as players sift through more documents, logs, and distractors. Another is \textit{progressive unlocking}, where later evidence or story beats become available only after players identify key violations, preventing downstream clues from implicitly revealing missed concepts. For example, in the Easy Healthcare case, some players who do not initially treat ``Log Fertility'' as health information disclosure may later infer the violation from evidence about failing to notify users, regulators, or the media. Removing these dependencies could further increase difficulty.

\section{Limitation and Future Work}

\sssec{Participant Recruitment.} We recruited students from a single U.S. university, with more undergraduate than graduate students, which may limit the generalizability of our findings to other institutions or populations. \revision{We treat our results as preliminary evidence of promising short-term gains rather than definitive proof of efficacy. Future work could recruit broader participant pools, including professional developers in industry settings, to confirm these gains and explore applications such as on-boarding training programs.}

\sssec{Lag Between Enforcement and Emerging Regulation.} Legal cases offer a unique learning opportunity by providing ``ground truth'' on how abstract privacy principles apply to real products. However, enforcement is slow and backward-looking. Complaints and orders often appear years after the conduct and focus on already-settled rules. \revision{As the enforcement corpus expands into emerging areas, the game's case library should evolve accordingly.}

\sssec{Jurisdictional Limitations.} We designed \name\ based on U.S. FTC settlements, reflecting U.S.-specific regulatory priorities. Privacy regulations differ across jurisdictions. For example, the EU's GDPR emphasizes data minimization, which is not required under any general federal privacy regulations in the U.S. Future work could extend this framework to incorporate privacy concepts emphasized in other jurisdictions’ regulations.

\sssec{Long-Term Effectiveness.} \revision{Measuring the long-term impact of educational game interventions remains an open challenge for the field~\cite{backlund2013educational, facchino2025use}. As a first step, our study examines whether \name yields meaningful short-term gains. Future work should examine whether these gains persist over time and translate into developers' real design decisions.}

%% file: parts/p10-ethical-considerations.tex
\section{Ethical Considerations}

This study received IRB approval and poses no more than minimal risk, comparable to typical computer-based coursework. 
Potential discomforts include mental fatigue, frustration with time-limited tasks, mild stress about test performance, and discomfort reflecting on privacy issues in real products. Participants could gain privacy awareness and reasoning skills that are rarely taught explicitly in standard curricula. All participants received a \$20 Amazon gift card as compensation for their time. Participants provided informed consent before beginning and could withdraw at any time without penalty. Overall, we believe the expected benefits outweigh the minimal risks to participants.

%% file: parts/p11-conclusion.tex
\section{Conclusion}

We introduce \name, a narrative investigation game that trains student developers' privacy awareness by examining real-world data practices derived from FTC enforcement cases. \revision{In a between-subjects study with 36 participants, we compared the game to a reading-based condition using the same FTC enforcement documents.} 
Participants who played \name showed significant gains in identifying violations, with higher recall and higher precision, and produced more complete justifications grounded in relevant evidence. In contrast, participants in the reading condition showed no significant pre-post improvements on these measures.

%% file: parts/p99-appendix.tex
\appendices

\section{Example Narrative Used in Pretests \& Posttests}
\label{app:narratives}

\begin{framed}
\small
Easy Healthcare Corporation develops Premom, a free ovulation and period-tracking app. The app encourages people trying to conceive to log health information such as menstrual cycles, fertility status, hormone test results, and pregnancy details. It collects health data from multiple sources, including user-entered information, uploaded photos of ovulation test strips, and body temperature data imported from Apple Health and Bluetooth-connected thermometers. When users connect a Bluetooth thermometer, Premom displays prompts instructing them to enable GPS and grant location access so the app can find and pair with the device.

Premom’s website and in-app privacy policies state that Easy Healthcare will not share users’ exact age or any data related to their health with third parties without the user’s consent or knowledge. The policies further state that the company collects only non-identifiable information for analytics, that third-party analytics and SDKs identify users solely by IP address, and that Premom uses data only to customize, measure, and analyze its services, content, and advertising.

In practice, Easy Healthcare integrated third-party SDKs, including Google Analytics, AppsFlyer, U-Share, and JPush, to track user interactions. Premom recorded user actions as Custom App Events with titles such as ``Calendar/Report/LogFertility'' and ``Log period-save'' and transmitted these events in plain text to Google and AppsFlyer. Through U-Share and JPush, the app transmitted users’ social media account information, precise GPS coordinates, and device identifiers (e.g., Android ID, IMEI, and MAC address). The privacy policies of U-Share and JPush state that they may use this data for advertising and share it with their partners. Easy Healthcare reviewed and accepted these terms. Easy Healthcare did not notify users, the Federal Trade Commission, or media outlets that it disclosed users’ health-related data and identifiers to third parties.
\end{framed}

\sssec{Violations \& Key Evidence (Hidden from participants)}
\begin{itemize}[noitemsep,leftmargin=*]
    \item Misrepresentation of Practices: promise not to share users’ health data, shared through third-party SDKs
    \item Misrepresentation of Practices: promise third parties collect only non-identifiable info, shared identifiable data
    \item Misrepresentation of Practices: claim to use data only to customize, measure, and analyze its own services, third parties can use the data for their own purposes
    \item Failure to Disclose Practices: failed to disclose that they shared location data
    \item Health Breach Notification Failure: breached health information, did not notify required parties
\end{itemize}

\section{List of Violation Types}
\label{app:violation-type-list}

\noindent See Table~\ref{tab:violation-categories}.

\begin{table}[h!]
\centering
\footnotesize
\setlength{\tabcolsep}{4pt}
\caption{List of violations used in pretest and posttest.}
\begin{tabularx}{\columnwidth}{@{}p{0.32\columnwidth}X@{}}
\toprule
\textbf{Violation Category} & \textbf{Description} \\
\midrule
Misrepresentation of Practices        & Making a consumer-facing claim inconsistent with the company's actual practices. \\
\midrule
Failure to Disclose Practices         & Omitting or incompletely disclosing information about the company's products or data practices. \\
\midrule
Health Breach Notification Failure    & Missing, delayed, or incomplete notice after a breach of health information. \\
\midrule
Excessive Retention of Children's Data & Keeping children's personal information longer than reasonably needed for the collection purpose. \\
\midrule
Failure of Parental Control           & Failing to provide parents effective means to review, delete, or limit collection or use of children's data. \\
\midrule
Failure of Parental Notice            & Failing to provide clear, complete, and accurate notice to parents about children's data practices. \\
\midrule
Failure to Obtain Verifiable Parental Consent & Collecting, using, or disclosing children's data without first obtaining verifiable parental consent. \\
\midrule
Failure to Clearly Disclose Transaction & Failing to clearly disclose all material terms before obtaining billing information, including negative option features. \\
\midrule
Failure to Obtain Express Informed Consent & Charging under a negative option feature without first obtaining the consumer's express informed consent. \\
\midrule
Failure to Provide Simple Cancellation & Failing to provide a simple, reasonable way for consumers to stop recurring charges. \\
\bottomrule
\end{tabularx}
\label{tab:violation-categories}
\end{table}

\section{Violation Templates}
\label{app:violationtemplates}

\noindent
See Table~\ref{tab:principle-schema}.

\begin{table*}[h!]
\centering
\footnotesize
\setlength{\tabcolsep}{4pt}
\renewcommand{\arraystretch}{0.95}
\caption{Violation templates for commonly cited privacy principles in FTC enforcement cases}
\begin{tabularx}{\linewidth}{@{} p{0.12\linewidth} l X @{}}
\toprule
\textbf{Violation} & \textbf{Field} & \textbf{Field definition} \\
\midrule
\multirow{2}{=}{Misrepresentation of Practices}
  & Claim
  & The consumer-facing statement, promise, or impression created by the company \\
\cmidrule(l){2-3}
& Actual Practice
  & The company's actual practice that differs from the claim \\
\midrule
\multirow{2}{=}{Failure to Disclose Practices}
  & Deficient Disclosure
  & The incomplete, missing, or obscured disclosure that fails to fully describe the company's practices \\
\cmidrule(l){2-3}
& Undisclosed Practice
  & The practice that is done but not fully or clearly disclosed \\
\midrule
\multirow{2}{=}{Health Breach Notification Failure}
  & Unauthorized Disclosure
  & Unauthorized disclosure or breach of personal health record information \\
\cmidrule(l){2-3}
& Notification Failure
  & Failure to notify affected individuals, the FTC, or media (for breaches affecting 500+ individuals) \\
\midrule
\multirow{3}{=}{Excessive Retention of Children's Data}
  & Child User Base
  & Evidence that the product or service collects personal information from children under 13 \\
\cmidrule(l){2-3}
& Data Collection from Children
  & Evidence that the operator collects personal information from those children \\
\cmidrule(l){2-3}
& Retention Duration
  & Actual retention period for children's personal information \\
\midrule
\multirow{2}{=}{Failure of Parental Control over Children’s Data}
  & Child User Base
  & Evidence that the product or service collects personal information from children under 13 \\
\cmidrule(l){2-3}
& Parental Control Failure
  & Failure to respect parental control rights to control their children's personal information \\
\midrule
\multirow{2}{=}{Failure of Parental Notice about Children's Data}
  & Child User Base
  & Evidence that the product or service collects personal information from children under 13 \\
\cmidrule(l){2-3}
& Parental Notice Omission
  & Failure to provide complete and truthful notice on the children’s data practice in the parental notice \\
\midrule
\multirow{2}{=}{Failure to Clearly Disclose Transaction Terms}
  & Deficient Disclosure
  & The incomplete, obscured, or unclear disclosure of material transaction terms before collecting billing information \\
\cmidrule(l){2-3}
  & Negative Option
  & The design or feature that treats consumer silence or inaction as agreement to be enrolled in, or continue, a paid transaction with recurring charges \\
\midrule
\multirow{2}{=}{Failure to Obtain Express Informed Consent}
  & Lack of Express Consent
  & How the company failed to obtain the consumer's clear, affirmative agreement to the negative option terms \\
\cmidrule(l){2-3}
  & Negative Option Charge
  & A charge resulted from treating silence or inaction as agreement to start or continue a paid subscription \\
\midrule
\multirow{2}{=}{Failure to Provide Simple Cancellation Mechanism}
  & Recurring Charge
  & Recurring payment that continues to be billed to the consumer under the negative option feature \\
\cmidrule(l){2-3}
  & Cancellation Mechanism Defect
  & How the company's cancellation process is missing, hidden, overly burdensome, or otherwise not simple for consumers to stop the recurring charges \\
\bottomrule
\end{tabularx}
\label{tab:principle-schema}
\end{table*}

\section{\revision{Residual Diagnostics for Mixed-Effects Models}}
\label{app:qq-plots}
\begin{figure}[h]
    \centering
    \includegraphics[width=\linewidth]{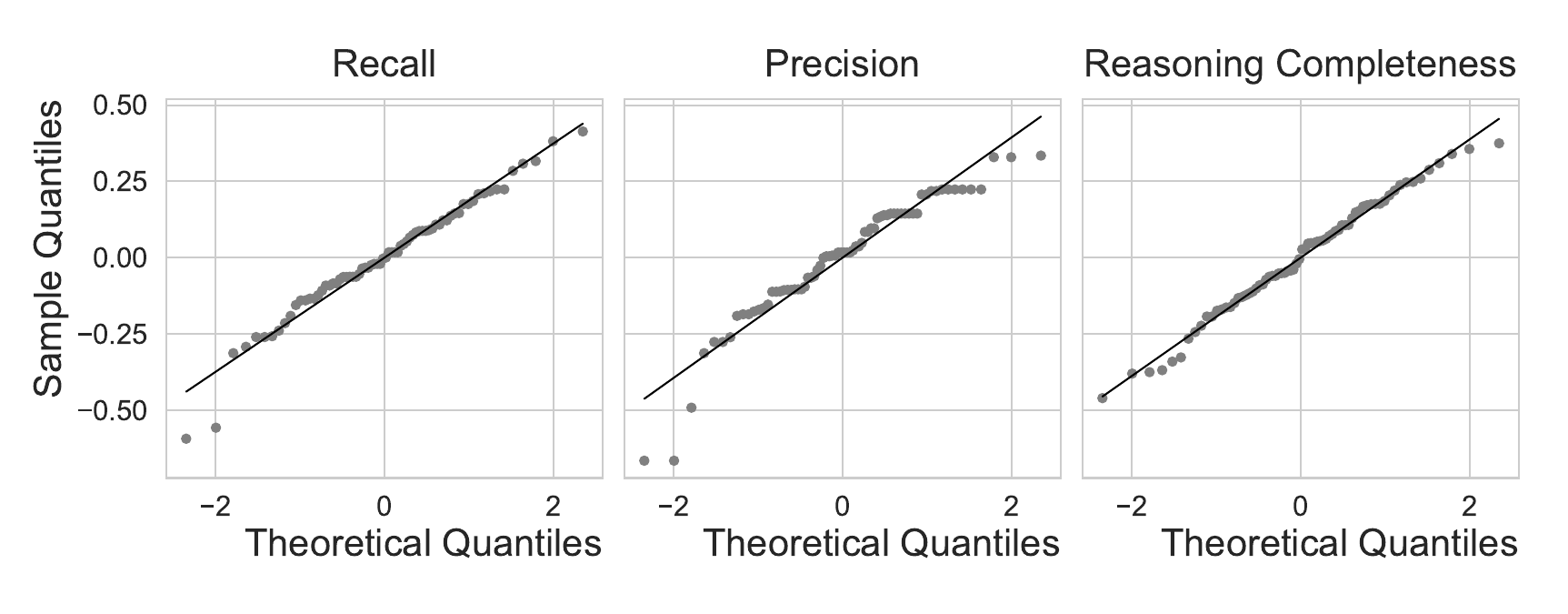}
    \caption{Q-Q plots of residuals from the three mixed-effects models. Points near the diagonal indicate that the residuals are approximately normally distributed. 
    }
    \label{fig:qq-plots}
\end{figure}

\noindent
See Figure~\ref{fig:qq-plots}.

\section{Qualitative Analysis Codebook}
\label{app:qualitative-codebook}

See Table~\ref{tab:qualitative-codebook}.

\begin{table*}[h!]
\footnotesize
\setlength{\tabcolsep}{3pt}
\centering
\caption{Qualitative analysis codebook.}
\label{tab:qualitative-codebook}
\begin{tabular}{p{0.11\linewidth} p{0.15\linewidth} p{0.28\linewidth} p{0.38\linewidth}}
\toprule
\textbf{Main Code} & \textbf{Main Code Definition} & \textbf{Subcode} & \textbf{Subcode Definition} \\
\midrule
\multirow{3}{=}{MC1: Intuition} &
\multirow{3}{=}{Participants' general intuition/reactions about the scenario or violations based on personal feelings} &
SC1.1: Gut feelings &
Participants found violations based on their personal gut feelings \\
\cmidrule(lr){3-4}
& & SC1.2: Perceived Irrelevance &
Belief that certain violations don't apply to the scenario at hand \\
\cmidrule(lr){3-4}
& & SC1.3: Prior Knowledge/Context &
Through previous experiences, the participant is able to relate...hat they are familiar with(ex: Company products, policies, etc.) \\
\midrule
\multirow{3}{=}{MC2: Learning and Confidence Progression} &
\multirow{3}{=}{Changes in ability/strategy to identifying violations from pretest to posttest} &
SC2.1: Increased Confidence + Ability to identify &
Greater certainty in violation identification after learning \\
\cmidrule(lr){3-4}
& & SC2.2: Increased Attention to Data Practice Details &
Participants demonstrate heightened attention to specific data practices in response to contextual cues (e.g., child-related data) \\
\cmidrule(lr){3-4}
& & SC2.3: Improved Precision &
Participants had reduced overgeneralization and improved precision in categorization \\
\midrule
\multirow{2}{=}{MC3: Confusion About Violations} &
\multirow{2}{=}{Participants had a hard time distinguishing multiple issues} &
SC3.1 : Overgeneralization &
Applying a single violation label overly broadly \\
\cmidrule(lr){3-4}
& & SC3.2: Multi-Category Confusion &
Uncertainty about which category applies when multiple seem relevant \\
\midrule
\multirow{2}{=}{MC4: Information Gaps and Needs} &
\multirow{2}{=}{Participant perception of missing information that would help them make judgments} &
SC4.1: Need for Complete Context &
Participants considered potential violations but rejected them due to lack of evidence \\
\cmidrule(lr){3-4}
& & SC4.2: Vagueness Created Confusion &
Vague or incomplete narratives made it harder to accurately identify violations; created confusion or an overall lack of confidence \\
\bottomrule
\end{tabular}
\label{tab:qualitative-codebook}
\end{table*}

\section{Participant Demographics}
\label{demographics}
\noindent
See Figure~\ref{tab:demographics-breakdown}.

\begin{table}[h!]
\centering
\footnotesize
\renewcommand{\arraystretch}{0.9}
\caption{Participant demographics.}
\begin{tabular}{llll}
\toprule
& \textbf{Category} & \textbf{Game} & \textbf{Reading} \\
\midrule

\multirow[t]{4}{*}{Gender}
& Male & 10 (55.6\%) & 12 (66.7\%) \\
& Female & 6 (33.3\%) & 5 (27.8\%) \\
& Non-binary & 1 (5.6\%) & -- \\
& Prefer not to say & 1 (5.6\%) & 1 (5.6\%) \\
\midrule

\multirow[t]{2}{*}{Age}
& 18--24 & 16 (88.9\%) & 15 (83.3\%) \\
& 25--34 & 2 (11.1\%) & 3 (16.7\%) \\
\midrule

\multirow[t]{3}{*}{Education}
& Undergraduate & 10 (55.6\%) & 10 (55.6\%) \\
& Graduate & 8 (44.4\%) & 8 (44.4\%) \\
\midrule

\multirow[t]{5}{*}{Major}
& Computer Science & 10 (55.6\%) & 13 (72.2\%) \\
& Data Science & 4 (22.2\%) & 2 (11.1\%) \\
& Elec. \& Comp. Eng. & 2 (11.1\%) & 1 (5.6\%) \\
& Bioinformatics & 1 (5.6\%) & 1 (5.6\%) \\
& Mathematics & 1 (5.6\%) & -- \\
& Cognitive Science & -- & 1 (5.6\%) \\
\midrule

\multirow[t]{4}{*}{Prog. Exp.}
& More than 5 years & 4 (22.2\%) & 4 (22.2\%) \\
& 4--5 years & 8 (44.4\%) & 4 (22.2\%) \\
& 3--4 years & 4 (22.2\%) & 5 (27.8\%) \\
& 2--3 years & 2 (11.1\%) & 5 (27.8\%) \\
\bottomrule
\end{tabular}
\label{tab:demographics-breakdown}
\end{table}